# Dynamical simulation of chiral induced spin-polarization and magnetization

*Peng-Yi Liu[1#], Tian-Yi Zhang[1#], and Qing-Feng Sun[1,2,*].*

[1]International Center for Quantum Materials, School of Physics, Peking University, Beijing, 100871, China.

[2]Hefei National Laboratory, Hefei 230088, China.

*Email: sunqf@pku.edu.cn

ABSTRACT. Despite generally lacking ferromagnetic properties or strong spin-orbit coupling, electrons in chiral molecules exhibit unique spin-dependent transport behavior, known as chiral-induced spin selectivity (CISS). This phenomenon implies a profound connection between chirality and spin, and draws attention to the link between chirality and magnetism. Experiments in recent years have shown that chirality can induce spin-polarizations and magnetizations, providing fresh insights into interpreting chirality-related biochemical processes and designing nano-magnetic devices. In this paper, we present a dynamical theoretical model aimed at elucidating how charge-polarization combined with the CISS leads to spin-polarization and magnetization. Our theoretical model successfully explains the spin-polarization and magnetization observed in three types of experiments, where the charge-polarization is induced in the chiral molecules by the dispersion interaction, gate voltage, and molecular adsorption. The model simulates a clear time evolution process and provides a comprehensive theoretical framework for this field.

Electron transfer in chiral materials exhibits a unique spin dependence, known as chiral-induced spin selectivity (CISS).[1,2] This phenomenon is particularly intriguing because it manifests prominently in organic chiral molecules that typically lack ferromagnetism or strong spin-orbit coupling (SOC), defying intuitive expectations. CISS uncovers a profound connection between chirality and spin, two fundamental properties in nature. It not only offers a novel perspective for understanding the intrinsic differences in enantiomers[3] but also exhibits broad application prospects in physics,[4–6] chemistry,[7,8] and biology.[9–11] Experimental studies[12–15] and theoretical analyses[16–21] collectively advance the understanding of CISS, establishing it as a significant theme at the intersection of chiral science and spintronics.

Spin selection is often closely associated with magnetism, leading to the inclination to link chirality with magnetism through CISS. In fact, many experiments in recent years have observed that chiral molecules [alpha-helix polyalanine (AHPA), for example] can induce spin-polarization and magnetization, which we categorize into three types. (i) Homochiral molecules exhibit stronger binding than heterochiral molecules[22] (with experimental understandings shown in the upper panel of Figures 1a, b, and Figure S1a). (ii) Anomalous Hall effect (AHE) as a signature of magnetization appears when a gate voltage is applied to chiral molecule/2DEG heterostructures[23–25] (upper panel of Figure 2a and Figure S1b). (iii) Chiral molecules can twist the magnetization direction of the substrate[26–29] (Figure 3a and Figure S1c) and the latter controls the adsorption rate of chiral molecules.[9,30] Despite the existence of partial theoretical understandings,[31–34] these results urgently necessitate a comprehensive theoretical model that encapsulates all observed phenomena, in order to advance the understanding of interactions between chiral molecules, achieve efficient separations of enantiomers, and apply molecular magnetism and spintronics at the nanoscale.

In this letter, we employ the Lindblad-type quantum master equation to establish a dynamical model of chiral induced spin-polarization and magnetization (CISM), which indicates that the combination of the charge-polarization and CISS can naturally lead to spin-polarization and magnetization. Based on this model, all three types of experimental phenomena mentioned above can be well understood, establishing a comprehensive and reliable theory in this field. Meanwhile, the use of dynamical equations allows for a clear depiction of the time evolution of spin-polarization.

In the experiments of CISM, AHPA is one of the most commonly used chiral materials, [2,22,26–28,35] with its two enantiomers denoted as AHPA-L and AHPA-D. The structure of the AHPA is illustrated in Figure S1, where each AHPA adopts a single-helical structure and exhibits CISS. To model AHPAs, we employ a tight-binding Hamiltonian $H_{AHPA}$ that has been reported,[17] describing a single-helical molecule with intrinsic chirality and weak SOC:

$$H_{AHPA} = \sum_{m=1}^{L-1} \sum_{j=1}^{L-m} \left( t_j c_m^\dagger c_{m+j} + 2\mathrm{i} t_j^S \cos \varphi_{mj}^- \, c_m^\dagger s_{mj} c_{m+j} + h.c. \right) \tag{1}$$

where $c_m^\dagger = [c_{m,\uparrow}^\dagger, c_{m,\downarrow}^\dagger]$ to represent the creation operator of electrons at site $m$ of an AHPA. $t_j$ is the hopping integral between site $m$ and site $m + j$, and $t_j^S$ is the corresponding SOC. $s_{mj}$ is a $2 \times 2$ matrix related to molecular structure. $\varphi_{mj}^- = j\Delta\varphi/2$, where $\Delta\varphi$ is the twist angle between two nearest sites, determining the helicity and chirality. $H_{AHPA}$ can successfully capture the key features of CISS[17] and the detailed expressions of the coefficients are provided in Supporting Information. The length of AHPAs is $L = 30$ sites. The twist angle is set to $\Delta\varphi = 5\pi/9$ for AHPA-Ls, $\Delta\varphi = -5\pi/9$ for AHPA-Ds, and achiral molecules are simulated by $\Delta\varphi = 0$. The right-to-left direction of the schematic diagrams in Figures 1-3 is marked as the $z$-direction, which is also the orientation of spin-↑. Other parameters we used are provided in Supporting Information.

Furthermore, we employ the Lindblad equation[36] to perform dynamical simulations of the density matrix[37]:

$$\hbar \frac{\mathrm{d}\rho}{\mathrm{d}\tau} = -\mathrm{i}[H, \rho] + \sum_{m=1} \sum_{\mu=x,y,z} \Gamma_m \left( L_{m\mu} \rho L_{m\mu}^\dagger - \frac{1}{2} \{ L_{m\mu}^\dagger L_{m\mu}, \rho \} \right) \tag{2}$$

where $\tau$ is the time, $\rho$ is the density matrix, $H$ is the Hamiltonian of the system, $L_{m\mu} = c_m^\dagger s_\mu c_m$ is the jump operator and $\Gamma_m$ is the jumping strength (details are provided in Supporting Information). After solving the density matrix $\rho(\tau)$, the spin-resolved electron density $n_m^{\uparrow/\downarrow}$ at site $m$ and the local spin-polarization $P_m = (n_m^\uparrow - n_m^\downarrow)/(n_m^\uparrow + n_m^\downarrow)$ can be obtained straightforwardly. In the following, we systematically introduce the physical modeling of the three types of experimental setups, present numerical results, and provide discussions on the findings.

The first type of experiments seeks to understand how the interaction between two homochiral molecules differs from the interaction between two heterochiral molecules.[22] The experimental scheme is to bring two chiral molecules close to each other and measure the interaction force between molecules with the same and opposite chiralities by the atomic force microscopes. The experiments found that the homochiral AHPAs have stronger binding than heterochiral AHPAs.[22] As shown in Figure S1a, previous numerical results showed that if two molecules are close to each other with the polarized spins forming a low-spin (high-spin) state, the energy is lower (higher) and the binding of two molecules is stronger (weaker).[22,23] Our subsequent results show how AHPAs with the same (opposite) chiralities give the low-spin (high-spin) state.

To theoretically simulate the experiment,[22] we construct a tight-binding model in which two AHPAs approach each other. Initially (the time $\tau \leq 0$), the two molecules are spatially separated and do not interact, each being described independently by the Hamiltonian $H_{AHPA}$ and remaining in equilibrium, and the whole Hamiltonian is $H(\tau < 0) = H_{AHPA} \oplus H_{AHPA}$. For the time $\tau > 0$, the two AHPAs begin to interact, as illustrated in the upper panels of Figures 1a (homochiral AHPAs) and 1b (heterochiral AHPAs). The dispersion interaction leads to charge redistribution, resulting in the formation of two parallel electric dipoles.[38] For simplicity, we introduce an effective electric field that captures the key features of dispersion-induced charge-polarization. In this field, the two AHPAs experience a potential energy $H_E = \sum_{m=1}^{2L}[\frac{2(m-1)}{2L-1} - 1]Vc_m^{\dagger}c_m$, with $V$ represents the strength of the electric field, which varies linearly along the molecular axis, while direct hopping of electrons between the AHPAs still does not exist. The whole Hamiltonian thus reads $H(\tau \geq 0) = H(\tau < 0) + H_E$. Based on this framework and the technical details provided in Supporting Information, we perform the dynamical simulations, with the results presented in Figures 1, S2, and S3.

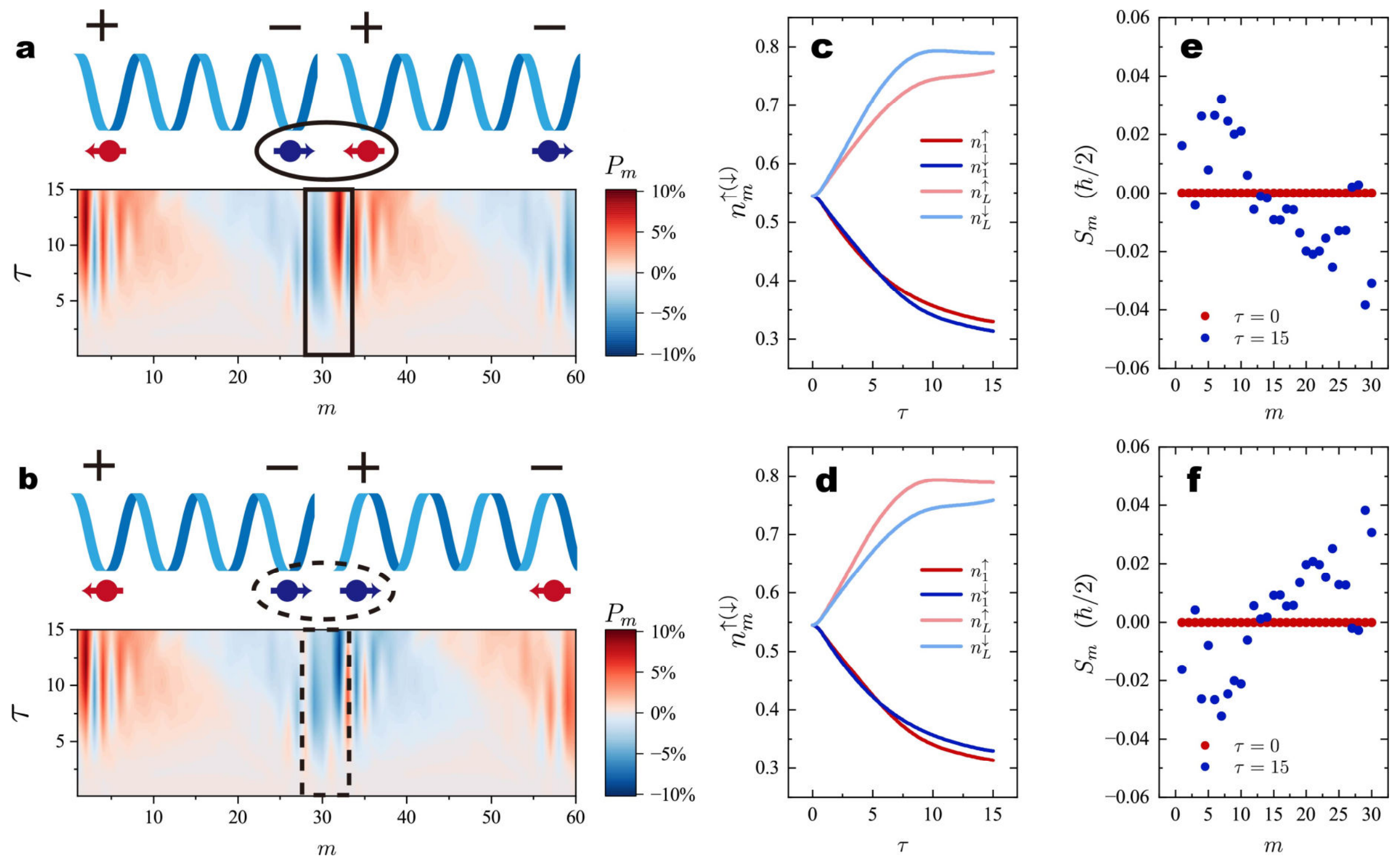

**Figure 1. Spin-polarization driven by the dispersion interaction.** (a) Upper panel: The schematic of two AHPA-Ls in proximity. "+" and "-" indicate charge-polarization, while red and blue arrows denote spin-↑ and spin-↓ polarization, respectively. The solid black box around $m = L = 30$ highlights a low-spin state

at their interface. Lower panel: Time evolution of local spin-polarization $P_m$, corresponding to the upper schematic. The left (right) AHPA contains sites with $1 \leq m \leq L$ ($L < m \leq 2L$). Red and blue label spin-↑ and spin-↓ polarized regions, respectively. (b) Upper panel: Schematic of an AHPA-L and an AHPA-D in proximity. The dashed black box highlights a high-spin state at their interface. Lower panel: Time evolution of local spin-polarization $P_m$, corresponding to the upper schematic. Symbols in (b) are the same as in (a). (c, d) The spin-resolved electron density at the first and the last sites are represented by $n_1^{\uparrow/\downarrow}$ and $n_L^{\uparrow/\downarrow}$ for an AHPA-L (c) and AHPA-D (d). (e) and (f) shows the local spin accumulation $S_m$ of an AHPA-L and AHPA-D, respectively, when $\tau = 0$ and $\tau = 15$.

The lower panel of Figure 1a shows the time evolution of the local spin-polarization $P_m$ when two homochiral molecules (two AHPA-Ls) are brought close together at the time $\tau = 0$. Before doing so, each AHPA has no spin-polarization with $P_m = 0$, due to the closed-shell electronic structure of the AHPA and the preservation of the time-reversal symmetry. But while the two AHPAs are in proximity, the dispersion force induces the electrons to move towards the same side of the two AHPAs,[22,23] leading to the formation of two electric dipoles in the same direction (as shown in Figures 1a, S1a, and S2a). In this process, as a result of CISS,[2,17] electrons with spin-↓ are more easily induced to move to one side, while electrons with spin-↑ are harder to drive and more remain in place. As shown in Figure 1a, opposite spin-polarization occurs on both sides of each AHPA-L. The spin-polarizations of the two AHPA-Ls are the same, so the opposite local spin-polarization occurs at the interface of the two AHPA-Ls (the solid black box around $m = L$ in Figure 1a), which becomes a low-spin and low-energy state.[22] Similarly, the proximity of heterochiral molecules also results in charge-polarization (Figure S2b) and spin-polarization $P_m$ (Figure 1b). However, due to the opposite chirality, the spin-polarizations of AHPA-L and AHPA-D are opposite. So, the local spin-polarization with the same direction occurs at the interface (the dashed black box in Figure 1b), which becomes a high-spin and high-energy state.[22]

Specifically, we focus on the time evolution of the spin-resolved electron densities of the first ($m = 1$) and the last ($m = L$) site of each AHPA, which are respectively represented by $n_1^\alpha(\tau)$ and $n_L^\alpha(\tau)$, $\alpha =\uparrow$ or $\downarrow$. For the AHPA-L, as shown in Figure 1c, when $\tau = 0$, we have $n_1^\uparrow = n_1^\downarrow = n_L^\uparrow = n_L^\downarrow$, indicating the absence of both charge-polarization and spin-polarization in the initial state. During time evolution, $n_1^\alpha$ decreases and $n_L^\alpha$ increases, demonstrating the emergence of charge-polarization. More importantly, at $\tau = 15$ in Figure 1c, $n_1^\uparrow > n_1^\downarrow$ ($n_L^\uparrow < n_L^\downarrow$) indicating the establishment of spin-$\uparrow$ (spin-$\downarrow$) polarization at $m = 1$ ($m = L$) of the AHPA-L. For AHPA-D, as depicted in Figure 1d, a same charge-polarization ($n_L^\alpha > n_1^\alpha$) and an opposite spin-polarization ($n_1^\uparrow < n_1^\downarrow$ and $n_L^\uparrow > n_L^\downarrow$) are observed, due to the different chiralities. We can also detect the local spin accumulation $S_m$ [$S_m \equiv \frac{\hbar}{2}\left( n_m^\uparrow - n_m^\downarrow\right)$, $\hbar$ is the reduced Plank constant] at each site to observe the spin-polarization, as shown in Figures 1e (AHPA-L) and 1f (AHPA-D). When $\tau = 0$, $S_m = 0$ is exactly for both the AHPA-L and AHPA-D. When $\tau = 15$, as shown in Figure 1e, the left region exhibits $S_m > 0$ while the right region shows $S_m < 0$, showing that AHPA-L develops spin-$\uparrow$ (spin-$\downarrow$) polarization on the left (right) side. In contrast, as shown in Figure 1f, AHPA-D exhibits an inverted spin-polarization pattern compared to AHPA-L.

We also establish an achiral control group in which the twist angle is set to $\Delta\varphi = 0$, thereby eliminating molecular chirality while keeping all other parameters unchanged. Although the charge-polarization still exists, as shown in Figure S3a, the densities of electrons with spin-$\uparrow$ and spin-$\downarrow$ are the same. Consequently, no local spin accumulation $S_m$ (Figure S3b) or local spin-polarization $P_m$ (Figure S3c) emerges at any site, highlighting the essential role of chirality.

From Figure 1, we can observe that when two AHPAs are close to each other, a low-spin (high-spin) state appears at the interface of two homochiral (heterochiral) AHPAs. According to the previous results of density functional theory,[22] our simulation shows that the binding of two homochiral molecules is stronger, and the binding of the heterochiral molecules is weaker. This is consistent with experimental results[22] and further provides an understanding of biometrics and homochirality of life.[11,39] In real systems, the charge-polarization of molecules is dynamic, and the magnitude or even direction of the electric dipole moment is not stable. However, it is not difficult to find from our results that as long as the two molecules have the electric dipole moments in the same direction, which is guaranteed by the dispersive interaction,[38] homochiral (heterochiral) molecules will still exhibit a low-spin (high-spin) state near the contact position.

The second type of experiments combined with the AHE created a new method for measuring CISS without a permanent magnet.[40] Specifically, the researchers assemble chiral molecules on the surface of a semiconductor heterojunction containing a 2DEG layer, fabricate a top gate to apply an electric field, and measure the Hall effect of the 2DEG under a longitudinal current,[23–25] as shown in the upper panel of Figure 2a and Figure S1b. When the gate voltage is 0, the system has no magnetism and no Hall voltage. After the gate voltage is turned on, the Hall voltage rapidly increases from 0 and then gradually decays back to 0. The two enantiomers correspond to opposite Hall voltages under the same gate voltages, whereas achiral molecules have almost no such effect. Experiments have attributed this AHE to electric field-driven charge-polarization accompanied by spin-polarization, causing 2DEG to experience magnetism. Our simulation results then show how the electric field drives the spin-polarization of chiral molecules and the CISM occurs.

Similar to the experimental setup, our model also considers an AHPA coupled to its substrate, which is described by a simple and non-chiral Hamiltonian $H_{sub} = \sum_{m=L+1}^{2L} \varepsilon_{sub} c_m^\dagger c_m + \sum_{m=L+1}^{2L-1} (t_{sub} c_m^\dagger c_{m+1} + h.c.)$. Here $c_m^\dagger = [c_{m,\uparrow}^\dagger, c_{m,\downarrow}^\dagger]$ is the creation operator at site $m$, $\varepsilon_{sub}$ is the on-site energy, and $t_{sub}$ is the hopping integral of the substrate. As shown in Figure 2a, the AHPA ($1 \leq m \leq L$) and the substrate ($L < m \leq 2L$) are located on the left and right sides, respectively.

Electron tunneling between the AHPA and the substrate is described by the coupling Hamiltonian $H_c = t_c c_L^\dagger c_{L+1} + h.c.$ where $t_c$ is the hopping integral. For $\tau < 0$, the system is governed by $H(\tau < 0) = H_{AHPA} + H_{sub} + H_c$ and remains in equilibrium. At $\tau = 0$, a gate voltage is turned on, which introduces an external electric field and linearly varied potential $H_E$ acting on both the AHPA and the substrate, as indicated by the black arrow in the upper panel of Figure 2a, thus $H(\tau \geq 0) = H_{AHPA} + H_{sub} + H_c + H_E$. Based on this framework, we can simulate the dynamical evolution of the electrons and spins for $\tau > 0$, with the results presented in Figures 2 and S4–S6. Additional details about the model are provided in Supporting Information.

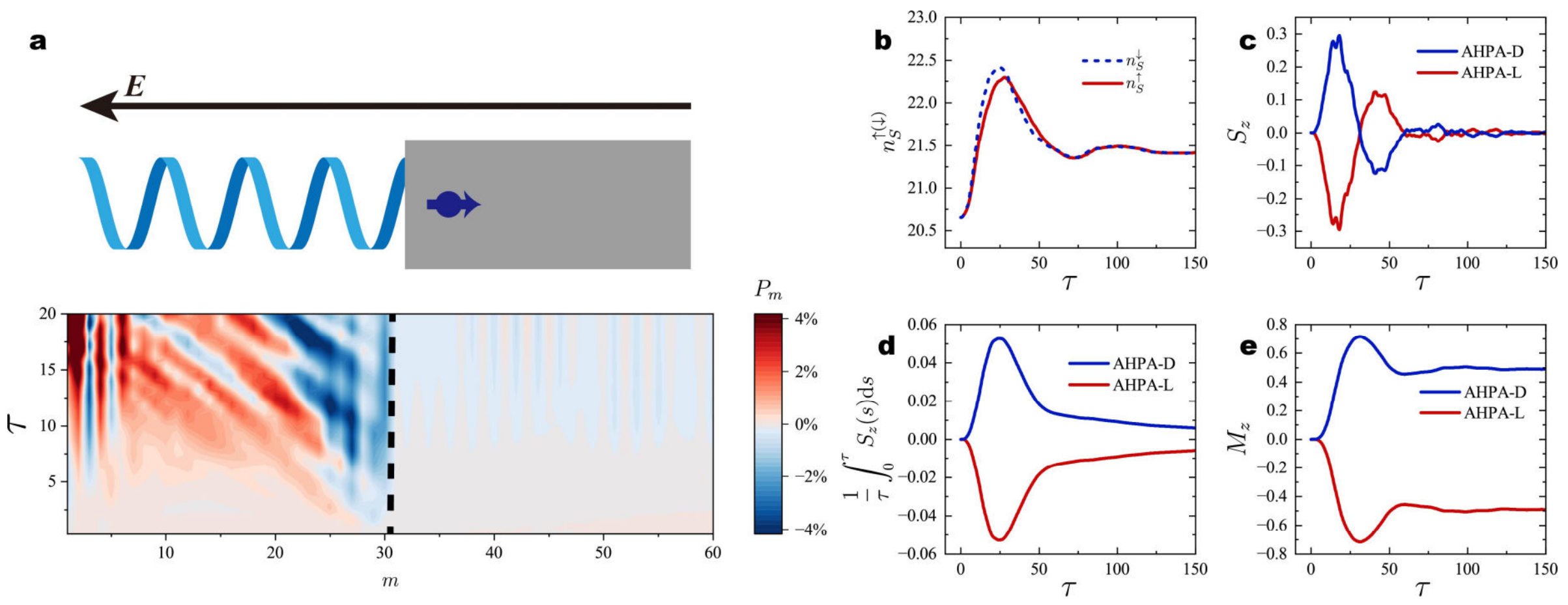

**Figure 2. Spin-polarization driven by a gate voltage.** (a) Upper panel: Schematic of an AHPA-L coupled to a substrate (the grey rectangle). The black arrow shows the electric field introduced by the gate voltage, and the small blue arrow in the substrate represents the spin-polarization. Lower panel: The time evolution of the local spin-polarization $P_m$ of the AHPA-L/substrate system, corresponding to the upper schematic. Red and blue show the spin-↑ and spin-↓ polarized regions, respectively. The dashed black line shows the interface of the AHPA-L and the substrate near $m = L$. (b) is the time evolution of $n_S^{\uparrow(\downarrow)}$ for the AHPA-L/substrate system, which reflects the spin-resolved number of electrons in the near-2DEG region ($2L/3 < m \leq 2L$). (c) and (d) respectively show the total spin accumulation $S_z$ of electrons and the time average of the total spin accumulation $\frac{1}{\tau}\int_0^\tau S_z(s)\,\mathrm{d}s$ in the near-2DEG region for two enantiomers, in the unit of $\hbar/2$. (e) shows the time evolution of the out-of-plane component of the magnetic moment $M_z$ in the unit of $|\boldsymbol{M}|$, when there is a magnetic moment in the substrate coupling to the spin accumulation.

As shown in the lower panel of Figure 2a, we study the time evolution of the local spin-polarization $P_m$ of the AHPA-L/substrate system under an applied electric field, corresponding to the upper schematic. When the time $\tau = 0$, no spin-polarization presents ($P_m = 0$). When $\tau > 0$, electrons begin to move as a result of the electric field, as shown in Figure S4a. In this process, electrons with spin-↓ are easier than those with spin-↑ to be pushed to the right by the electric field, due to CISS. As a result, a spin-↓ (spin-↑) polarization occurs on the right (left) side of the AHPA-L/substrate system (see Figure 2a). Similarly, when an AHPA-D is coupled to the substrate and turns on the gate voltage, a similar charge-polarization (Figure S4b) and opposite spin-polarization (Figure S5a) appear, and the latter is due to opposite chirality.

What really affects electron transport in the 2DEG is the spin-polarization close to the 2DEG, this may be possible through potential correlations such as RKKY interactions,[2,23] so we focus on the spin-polarization in the sites close to the substrate ($2L/3 < m \leq L$) and in the substrate ($L < m \leq 2L$) to ensure the possibility of the correlation. Specifically, we can evaluate the spin-resolved number of electrons in the near-2DEG region with $2L/3 < m \leq 2L$, represented by $n_S^{\uparrow(\downarrow)} = \sum_{m=2L/3+1}^{2L} n_m^{\uparrow(\downarrow)}$, and the total spin accumulation in this region $S_z = \frac{\hbar}{2}(n_S^{\uparrow} - n_S^{\downarrow})$. As shown in Figure 2b, the number of electrons $n_S^{\uparrow(\downarrow)}$ in the near-2DEG region first increases, then oscillates, and finally stabilizes at a value higher than the initial one, which is the process of charge-polarization. Importantly, in Figure 2b, CISS of the AHPA-L causes the number of electrons with spin-↓ to be obviously higher than the number of electrons in spin-↑ at the beginning of the charge transfer process, *i.e.*, a spin-↓ polarization presents in the near-2DEG region of the AHPA-L/substrate system. In contrast, the initial spin-polarization in this region caused by AHPA-D is ↑, as shown in Figure S5b.

The total spin accumulation $S_z$ in the near-2DEG region is shown in Figure 2c. Starting from the time $\tau = 0$, the absolute value $|S_z|$ rapidly increases due to the combined effects of the charge transfer and CISS. Subsequently, as a result of electron reflection and the end of charge transfer, $S_z$ gradually oscillates and decays to 0. $S_z$ exhibits a completely opposite dependence on the chirality of the molecule, leading to the experimental observation that opposite chiralities induce opposite Hall voltages.[23–25] Due to the capacitance effect and the correlation of accumulated spin, the time average of spin accumulation, rather than itself, is more likely to be related to the observed AHE, which increases and then decreases without changing signs, as shown in Figure 2d, which is consistent with the experiments.[23]

Then, we examine a control group in which achiral molecules were used instead of AHPAs. As shown in Figure S6, although electrons are driven by the gate voltage and forming a charge-polarization, all observable results become spin-independent, confirming that the gate voltage-induced spin-polarization and AHE are entirely due to CISS. The results were consistent with experimental observations,[23] where achiral molecules failed to induce the AHE.

Furthermore, since the magnetization response in the experiment could exceed 1 MHz without special optimization,[24] we further propose that a similar device can be used as a way to reverse the magnetic moment controlled by the electric field to achieve efficient and high spatial precision magnetic moment manipulation. It is considered that there is a magnetic moment $\boldsymbol{M}$ interacting with the spin accumulation in the region $2L/3 < m \leq 2L$, and the magnetic moment is initially magnetized in-plane. When the electric field is turned on, the spin accumulation acts as an effective magnetic field $\boldsymbol{H}_{\mathbf{eff}} = 2\alpha_M S_z \boldsymbol{e}_{\boldsymbol{z}}/\hbar|\boldsymbol{M}|$ through spin exchange and spin torque interaction,[31] and twists the magnetic moment out-of-plane, which can be described by the Lifshitz–Landau–Gilbert equation (see details in Supporting Information).[31,41] Here, $\alpha_M$ represents the interacting strength, $S_z$ is the accumulated spin defined above, and the unit vector $\boldsymbol{e_z}$ points to $+z$. The results are shown in Figure 2e, the out-of-plane magnetization component appears significantly under the action of the electric field. AHPA-L and AHPA-D can achieve opposite magnetization in the substrate, which can be stable and persistent over time.

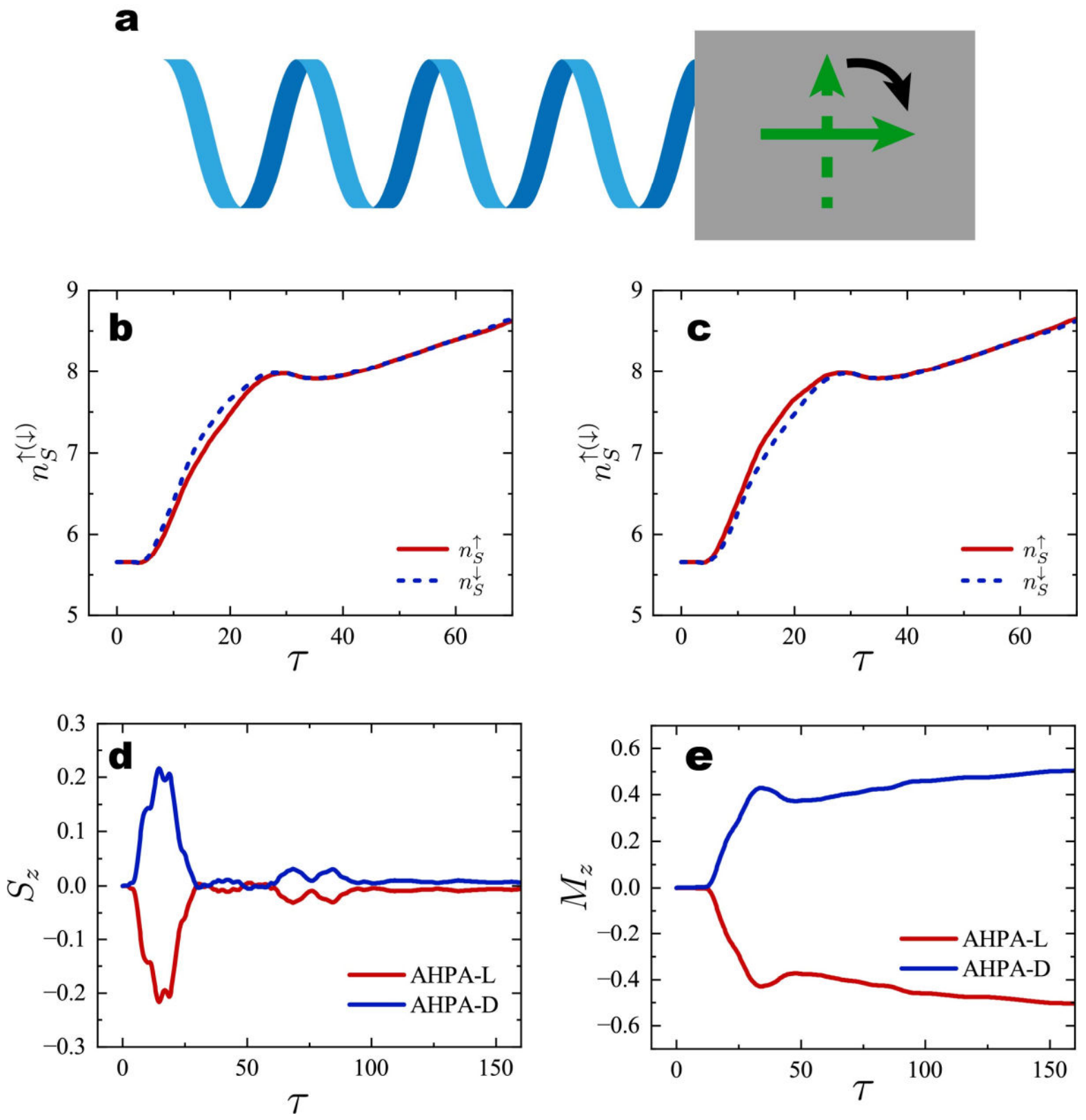

**Figure 3. Spin-polarization driven by the absorption.** (a) The schematic of an AHPA-L adsorbed on a magnetic substrate. The green arrows show that the magnetic moment in the substrate is twisted by molecular adsorption. (b) and (c) are the time evolution of the spin-resolved $n_S^{\uparrow(\downarrow)}$ in the region $2L/3 < m \leq 2L$, for the case of AHPA-L and AHPA-D, respectively. (d) shows the total spin accumulation $S_z$ in the region $2L/3 < m \leq 2L$ for two enantiomers, in the unit of $\hbar/2$. (e) is the time evolution of the out-of-plane component of the magnetic moment $M_z$, in the unit of $|\boldsymbol{M}|$.

The third type of experiments found that when chiral molecules are assembled on a magnetic substrate, the original magnetization direction of the substrate is twisted by the charge transfer,[26–28,42] as shown in Figure 3a and Figure S1c. This charge transfer arises because, upon adsorption of molecules onto the substrate, electrons redistribute to align the chemical potentials of the molecules and the substrate. For example, according to magnetic imaging by atomic force microscopy, AHPAs with one chirality can significantly enhance the magnetization of the out-of-plane component of the magnetic substrate, while the opposite chirality shows opposite magnetization due to the charge transfer and CISS.[26] Furthermore, spin-polarization and the interaction of magnetic substrates that occur during the adsorption have been applied to enantiomer separation.[9,30]

To understand this process, we consider the adsorption between an AHPA (with sites $1 \leq m \leq L$) and an achiral substrate (with sites $L < m \leq 2L$) with mismatched chemical potentials. For simplicity, at the time $\tau < 0$, there is no coupling between the substrate and the AHPA and all electrons reside in the AHPA, which remains in equilibrium under the Hamiltonian $H_{AHPA}$, while the states of the substrate with corresponding energies remain unoccupied. The whole Hamiltonian $H(\tau < 0) = H_{AHPA} + H_{sub}$. At the time $\tau = 0$, the molecule and substrate come sufficiently close to allow electron transfer (see Figure S1c), introducing the coupling Hamiltonian $H_c$, and $H(\tau \geq 0) = H_{AHPA} + H_{sub} + H_c$. Following this, the electron transfer from the AHPA to the substrate, driven by the trend in chemical potential matching, can be described by the Lindblad equation [see Equation (2) and Section S2 in Supporting Information].

We also use $n_S^{\uparrow(\downarrow)}$ to represent the spin-resolved electron number in the substrate and near-substrate with $2L/3 < m \leq 2L$. As shown in Figures 3b,c, both $n_S^{\uparrow}$ and $n_S^{\downarrow}$ increase over time regardless of chirality, indicating electron transfer from the left to the right to achieve chemical potential equilibrium. In this process, as shown in Figure 3b, due to the spin selectivity of AHPA-L, electrons with spin-↓ transfer faster than those with spin-↑, resulting in $n_S^{\downarrow} > n_S^{\uparrow}$ and a spin-↓ polarization in this region. For the case of AHPA-D, the opposite chirality leads to the opposite spin selectivity, so $n_S^{\downarrow} < n_S^{\uparrow}$ and a spin-↑ polarization occurs, as shown in Figure 3c. As the time $\tau$ increases, charge transfer slows down, accompanied by a weakening of spin-polarization for both the AHPA-L and AHPA-D. Then, we also examine the spin accumulation $S_z$ in the region $2L/3 < m \leq 2L$. As illustrated in Figure 3d, the total spin accumulation $S_z$ in this region rapidly increases over time, reaching a significant peak as a joint consequence of charge transfer and CISS. Subsequently, $S_z$ begins to oscillate and decay due to the gradual establishment of charge equilibrium. More importantly, we can clearly see that the AHPAs with opposite chirality lead to the opposite spin-polarization, which reflects the nature of CISS.

Furthermore, to simulate the persistent magnetization observed in the experiments, we take the same configuration as the experiments and set a magnetic moment $\boldsymbol{M}$ in the substrate, which is initially ($\tau < 0$) magnetized in-plane (the dashed green arrow in Figure 3a). When the spin accumulation $S_z$ appears close to the moment, the magnetic moment feels an effective magnetic field $\boldsymbol{H}_{\mathbf{eff}} = 2\alpha_M S_z \boldsymbol{e}_{\boldsymbol{z}}/\hbar|\boldsymbol{M}|$ mentioned above,[31] beginning to evolve in time according to the Lifshitz–Landau–Gilbert equation.[31,41] With the evolution of time, the magnetic component $M_z$ out of the plane experiences a rapid increase, gradually saturates, and finally stabilizes at a significant value $M_z \approx 0.5|\boldsymbol{M}|$, as shown in Figure 3e. Importantly, the AHPAs with the opposite chiralities induce the opposite out-of-plane magnetizations, as a result of the opposite spin selectivity, which is especially interesting and consistent with the results in the CISM experiment.[26]

In Figure 3e, we can see that the ferromagnetic substrate acts to stabilize the spin-polarization,[31] which makes the observed CISM in the experiment appear for a long time,[28] rather than a transient one.[23] This way of efficiently regulating magnetization through the adsorption process of molecules provides a new possibility for the design of magnetic devices.[26] In experiments, when the substrate is stabilized in out-of-plane magnetization by an external magnetic field, another phenomenon emerges: The adsorption of chiral molecules onto the ferromagnetic substrate is either promoted or hindered, depending on the direction of the out-of-plane magnetization of the ferromagnet and the chirality of the molecules.[9,30] Based on the above simulation, the adsorption process requires charge transfer, and the transferred electrons are spin-polarized. According to previous results, when the transferred spins are parallel (anti-parallel) to the substrate magnetization, the adsorption results in the formation of a high-spin state (low-spin state), which is relatively unstable (stable).[9,22,35] Consequently, the separation of enantiomers can be achieved by controlling the direction of substrate magnetization through a magnetic field.

As a control group, although the achiral molecule also experiences charge transfer during adsorption, neither spin-polarization nor out-of-plane magnetization in the substrate appears, as shown in Figure S7. This means that the CISM discussed above comes entirely from CISS, not from the magnetism of the open-shell molecule.[43,44] Further, we also investigate the influence of parameters on CISM, with the results shown in Figure S8. CISM remains observable for $\alpha_M$ over a wide range[31] and exhibits temperature robustness from zero to room temperature.

Finally, the influence of various parameters on CISM under more realistic conditions is analyzed in Sections S5 and S6 of the Supporting Information, including parameters of substrate, temperature and dephasing, and geometric parameters (Figure S9-S11).

In summary, we show that this chiral induced spin-polarization and magnetization effect can be attributed to the combination of CISS and charge-polarization. Particularly, the charge redistribution caused by the dispersion interaction of two molecules is accompanied by spin-polarization, which leads to the chirality-dependent binding of molecules. When the gate voltage drives the chiral molecule, spin-polarized electrons are injected into the substrate, which can result in the time-reversal-breaking AHE in the substrate. The spin-polarized charge transferred when the chiral molecule adsorbed on the ferromagnetic substrate is recorded as a stable magnetization in the substrate. All these experimental phenomena can be simulated and understood by the dynamic model. This work contributes to the application of molecular devices in spintronics and magnetism, as well as the understanding of the mechanism of CISS.

SUPPORTING INFORMATION

Detailed theoretical modeling: The Hamiltonian, dynamical simulation of electrons, dynamical simulation of magnetization, parameters we used, and parameter dependence of the model. Additional figures such as charge polarization and the control groups containing Figures S1-S11. (PDF)

AUTHOR INFORMATION

**Corresponding Author**

*Qing-Feng Sun* - International Center for Quantum Materials, School of Physics, Peking University, Beijing, 100871, China; Hefei National Laboratory, Hefei 230088, China. Email: sunqf@pku.edu.cn

**Author Contributions**

[#]P.-Y. L and T.-Y. Z contributed equally to this work.

**Notes**

The authors declare no competing financial interest.

ACKNOWLEDGMENT

This work was financially supported by the National Key R and D Program of China (Grant No. 2024YFA1409002), the National Natural Science Foundation of China (Grant No. 12374034), and the Innovation Program for Quantum Science and Technology (2021ZD0302403). The authors also acknowledge the High-performance Computing Platform of Peking University for providing computational resources.

# SUPPORTING INFORMATION

# Dynamical simulation of chiral induced spin-polarization and magnetization

*Peng-Yi Liu[1#], Tian-Yi Zhang[1#], and Qing-Feng Sun[1,2,]*.*

[1]International Center for Quantum Materials, School of Physics, Peking University, Beijing, 100871, China.

[2]Hefei National Laboratory, Hefei 230088, China.

[#]Contributed equally

*Email: sunqf@pku.edu.cn

The Supporting Information contains:

- Detailed theoretical modeling: The Hamiltonian, dynamical simulation of electrons, dynamical simulation of magnetization, and model parameters.
- Additional figures such as charge-polarization and the control groups (Figures S1-S11).
- References

**Contents**

# S1. Hamiltonian

The Hamiltonians used in the main text include the chiral molecule $H_{AHPA}$, substrate $H_{sub}$, coupling $H_c$, and electric field $H_E$. Below, we introduce these four parts one by one.

**(i)** The AHPA has a single-helical structure and is one of the most commonly used materials to achieve CISS.[1–5] In order to simulate the situation in real experiments, we use the tight-binding Hamiltonian $H_{AHPA}$ of a single-helical molecule, which can successfully simulate the CISS,[6] to study the processes of CISM.

$$\begin{aligned} H_{AHPA} &= H_{mol} + H_{SOC}, \\ H_{mol} &= \sum_{m=1}^{L-1}\sum_{j=1}^{L-m} t_j c_m^\dagger c_{m+j} + h.c., \\ H_{SOC} &= \sum_{m=1}^{L-1}\sum_{j=1}^{L-m} 2\mathrm{i}t_j^S \cos\varphi_{mj}^- \, c_m^\dagger s_{mj} c_{m+j} + h.c. \end{aligned} \tag{S1}$$

Here, $H_{mol}$ describes the Hamiltonian without SOC and $H_{SOC}$ is the SOC part. "$h.c.$" represents the Hermitian conjugate. We use $c_m^\dagger = [c_{m,\uparrow}^\dagger, c_{m,\downarrow}^\dagger]$ to represent the creation operator of electrons at site $m$ of an AHPA, where ↑ and ↓ represent the spin degree of freedom. The length of the AHPA is $L = 30$, and the potential energy without the external electric field is set to be zero without the loss of generality. The geometry of each AHPA[6] can be described by the radius $R$, the twist angle between the nearest sites $\Delta\varphi$, and the pitch of the helix $\Delta h$. Then, the Euclidean distance between sites $m$ and $m + j$ is $l_j = \sqrt{[2R\sin(j\Delta\varphi/2)]^2 + (j\Delta h)^2}$, and the corresponding space angle $\theta_j = \arccos[2R\sin(j\Delta\varphi/2)/l_j]$ . $t_j = t_1 e^{-(l_j - l_1)/l_c}$ is the long-range hopping energy[6,7] between site $m$ and site $m + j$. Similarly, $t_j^S = t_1^S e^{-(l_j - l_1)/l_c}$ is the corresponding SOC. $t_j$ and $t_j^S$ decays with the exponent $l_c$. According to our previous results,[6] $s_{mj} = (s_x \sin\varphi_{mj}^+ - s_y \cos\varphi_{mj}^+)\sin\theta_j + s_z\cos\theta_j$, with $\varphi_{mj}^\pm = [(m + j) \pm m]\Delta\varphi/2$. In the whole paper, $s_x$, $s_y$, and $s_z$ represent the Pauli matrices in the $x$, $y$, and $z$ directions, respectively.

**(ii)** The substrate is simulated by a simple achiral tight-binding chain containing $L$ sites, with Hamiltonian $H_{sub} = \sum_{m=L+1}^{2L} \varepsilon_{sub} c_m^\dagger c_m + \sum_{m=L+1}^{2L-1} (t_{sub} c_m^\dagger c_{m+1} + h.c.)$ with $c_m^\dagger = [c_{m,\uparrow}^\dagger, c_{m,\downarrow}^\dagger]$ is the creation operator of electrons at site $m$ of the substrate ($L < m \leq 2L$). The real substrate has a band that is much wider than the molecular energy levels. $H_{sub}$ we adopt here only simulate the substrate energy bands close to the molecular energy levels, because the energy bands far from the molecular energy levels do not have a significant impact on the processes involving molecules.

**(iii)** The coupling between the AHPA and the substrate is set to be spin-independent $H_c = t_c c_L^\dagger c_{L+1} + h.c.$.

**(iv)** If an electric field is turned on, the Hamiltonian will have a potential energy that varies linearly with position $H_E = \sum_{m=1}^{2L} [\frac{2(m-1)}{2L-1} - 1] V c_m^\dagger c_m$, and $V$ represents the strength of the electric field.

## S2. Dynamical simulation of electrons

The simulation of the charge and spin dynamics is performed by the Lindblad equation, which describes the evolution of the density matrix of electrons in an open quantum system.[8] The time evolution incorporates both unitary quantum mechanical dynamics and dephasing effects induced by the environment, the latter of which has been widely recognized as both ubiquitous and significant in organic systems and CISS.[1,9–11]

$$\hbar \frac{\mathrm{d}\rho}{\mathrm{d}\tau} = -\mathrm{i}[H, \rho] + \sum_{m=1}^{2L} \sum_{\mu=x,y,z} \Gamma_m \left( L_{m\mu} \rho L_{m\mu}^\dagger - \frac{1}{2} \{ L_{m\mu}^\dagger L_{m\mu}, \rho \} \right). \quad \text{(S2)}$$

Here, $\rho$ is the single-particle density matrix. $[\ldots, \ldots]$ and $\{\ldots, \ldots\}$ are the commutator and the anticommutator, respectively. The first term of Eq. (S2) is the unitary evolution of the quantum states, and the second term describes the dephasing effect of the environment. In experiments at room temperature, the dephasing effect is significantly present in organic[9,10] and inorganic[12,13] systems through inelastic scatterings from the electrons, impurities, nuclear spins, and phonons. As a result, electrons lose their phase and spin memory during the time evolution. The quantum jump operator is given by $L_{m\mu} = c_m^\dagger s_\mu c_m$. This kind of jump operator comes from the interaction between the system and the environment,[8] which can relax phase memory, relax spin memory, and keep the number of particles conserved. In Figure 1, $\Gamma_m = \Gamma_{AHPA}$ is a constant, corresponding to a uniform dephasing strength of two AHPAs. In Figure 2 and Figure 3, $\Gamma_m = [\Gamma_{AHPA}\theta(L + 1/2 - m) + \Gamma_{sub}\theta(m - L - 1/2)]$, where $\theta(\dots)$ is the step function. $\Gamma_{AHPA}$ and $\Gamma_{sub}$ are the dephasing rates in the AHPA and the substrate, respectively. Before studying time evolution, we need to determine the Hamiltonian and initial state of the system as follows.

**(i)** When two AHPA are close to each other, dispersion-induced charge re-organization occurs, which can be simulated by an effective electric field. As we mentioned in the main text, when the time $\tau \leq 0$, two AHPAs are isolated $H(\tau < 0) = H_{AHPA} \oplus H_{AHPA}$. The chirality of each AHPA is controlled by the parameter $\Delta\varphi$ in the Hamiltonian. Here, the direct sum "⊕" represents those two matrices of $H_{AHPA}$ are stacked diagonally, so there is no coupling between two AHPAs, and the Hilbert space dimension of $H(\tau < 0)$ becomes twice that of $H_{AHPA}$. When two AHPAs are close to each other, the dispersion interaction appears, so the electric potential is added to the Hamiltonian $H(\tau \geq 0) = H(\tau < 0) + H_E$. Since $H_E$ only shifts the potential energy, the direct hopping between the two molecules still does not exist. The initial state is given by the equilibrium state of $H(\tau < 0)$, $\rho(\tau < 0) = \rho_{AHPA} \oplus \rho_{AHPA} / \mathrm{tr}(\rho_{AHPA} \oplus \rho_{AHPA})$, where each molecular density matrix $\rho_{AHPA}$ is constructed in the eigenstate $|\psi_i\rangle$ of $H_{AHPA}$:

$$\rho_{AHPA} = \frac{1}{N} \sum_{i=1}^{2L} f(E_i) \, |\psi_i\rangle\langle\psi_i| \tag{S3}$$

where $E_i$ and $|\psi_i\rangle$ are the energy, $i$-th eigenstate of $H_{AHPA}$ : $H_{AHPA}|\psi_i\rangle = E_i|\psi_i\rangle$ . $f(E_i) = 1/\{\exp[(E_i - \mu_{mol})/k_B\mathcal{T}] + 1\}$ is the Fermi distribution, where $\mu_{mol}$ is the chemical potential of AHPA and set to $\mu_{mol} = 0$ in our study. The total number of electrons we care about is thus $N = \sum_i f(E_i) = \sum_i 1/[\exp(E_i/k_B\mathcal{T}) + 1] = 68$. Each AHPA has 34 electrons, slightly larger than the number of sites, because the long-range hopping breaks the electron-hole symmetry.[6]

**(ii)** When we discuss the effect of the gate voltage, we couple the AHPA and the substrate and turn on the electric field at $\tau = 0$. Then, $H(\tau < 0) = H_{AHPA} + H_{sub} + H_c$ and $H(\tau \geq 0) = H_{AHPA} + H_{sub} + H_c + H_E$. The initial density matrix is constructed in the eigenstate $|\psi_i\rangle$ of the full Hamiltonian $H(\tau < 0)$:

$$\rho(\tau < 0) = \frac{1}{N}\sum_{i=1} f(E_i)\,|\psi_i\rangle\langle\psi_i| \tag{S4}$$

where $H(\tau < 0)|\psi_i\rangle = E_i|\psi_i\rangle$, $f(E_i) = 1/[\exp(E_i/k_B\mathcal{T}) + 1]$ and $N = \sum_i f(E_i)$. Initially, there are 34 electrons in the AHPA, the same as in case (i), 30 electrons in the substrate, and $N = 64$.

**(iii)** When we discuss the effect of adsorption, we couple the AHPA and the substrate at $\tau = 0$. So, $H(\tau < 0) = H_{AHPA} + H_{sub}$ and $H(\tau \geq 0) = H_{AHPA} + H_{sub} + H_c$ . When $\tau < 0$ , we set electrons forming an equilibrium state of $H_{AHPA}$ with $N = 34$, and the substrate is empty. Since both the AHPA and substrate are in their own thermal equilibrium states, the density matrix is the normalized direct sum of their individual density matrices:

$$\rho(\tau < 0) = \frac{\rho_{AHPA} \oplus \sum_{i=1} f_{sub}(\varepsilon_i)\,|\phi_i\rangle\langle\phi_i|}{\mathrm{tr}(\rho_{AHPA} \oplus \sum_{i=1} f_{sub}(\varepsilon_i)\,|\phi_i\rangle\langle\phi_i|)} \tag{S5}$$

Here, $\rho_{AHPA} = 1/N\sum_{i=1}^{2L} f(E_i)\,|\psi_i\rangle\langle\psi_i|$ is constructed from the eigenstate $|\psi_i\rangle$ of $H_{AHPA}$ : $H_{AHPA}|\psi_i\rangle = E_i|\psi_i\rangle$, $f(E_i) = 1/[\exp(E_i/k_B\mathcal{T}) + 1]$ and $N = \sum_i f(E_i)$. The substrate density matrix is constructed from the eigenbasis $|\phi_i\rangle$ of $H_{sub}$ : $H_{sub}|\phi_i\rangle = \varepsilon_i|\phi_i\rangle$ , and $f_{sub}(\varepsilon_i) = 1/(\exp[(\varepsilon_i - \mu_{sub})/k_B\mathcal{T}] + 1)$. Because of the mismatch between the molecule and substrate

$\mu_{mol} \neq \mu_{sub}$ (for instance, we consider the case where $\mu_{mol} > \mu_{sub}$), substrate states described by $H_{sub}$ exhibit almost no electron occupation, so we set $f_{sub}(\varepsilon_i) = 0$ when $\tau < 0$. In this way, after the appearance of $H_c$, electrons will spontaneously diffuse from AHPA to the substrate.

Based on the above settings for the Hamiltonians and initial states at the time $\tau = 0$, we can obtain the density matrix $\rho(\tau)$ for the time $\tau > 0$ according to Eq. (S2) and calculate observables from the density matrix. The densities of electrons with spin ↑ and ↓ at site $m$ are given by $n_m^{\uparrow} = N\mathrm{tr}\big(\rho c_{m,\uparrow}^{\dagger} c_{m.\uparrow}\big)$ and $n_m^{\downarrow} = N\mathrm{tr}\big(\rho c_{m,\downarrow}^{\dagger} c_{m,\downarrow}\big)$, respectively. So, the local density of electrons and local spin accumulation are $n_m = n_m^{\uparrow} + n_m^{\downarrow}$ and $S_m = \frac{\hbar}{2}(\, n_m^{\uparrow} - n_m^{\downarrow})$, respectively. Finally, the local spin-polarization is $P_m = (\, n_m^{\uparrow} - n_m^{\downarrow})/(n_m^{\uparrow} + n_m^{\downarrow})$.

# S3. Dynamical simulation of magnetization

In Figure 2e (the end of the second experiment) and Figure 3 (the third experiment), we study the influence of chiral molecules on the magnetization in the substrate. It is well known that the magnetic moment can be twisted under the action of an external magnetic field.[14] Additionally, mechanisms such as spin-exchange interaction and spin torque allow polarized spins to influence nearby magnetic moments, as suggested by a recent CISM study.[15] Specifically, we set that when spins in the AHPA are sufficiently close to the substrate (within approximately 1 nm, for $2L/3 < m \leq L$) or enter the substrate ($L < m \leq 2L$), they can significantly twist the magnetic moment $\boldsymbol{M}$ in the substrate, as such an effect becomes possible at a short distance. Initially, the magnetic moment is polarized in-plane. The specific orientation of the in-plane polarization is not important, due to the average of magnetic domains.[4,15] The time evolution of the magnetic moment is determined by the Lifshitz–Landau–Gilbert equation.[14,15]

$$\frac{\mathrm{d}\boldsymbol{M}}{\mathrm{d}\tau} = -\gamma \boldsymbol{M} \times \boldsymbol{H}_{\mathbf{eff}} - \lambda \boldsymbol{M} \times (\boldsymbol{M} \times \boldsymbol{H}_{\mathbf{eff}})\,. \tag{S5}$$

Here, $\gamma$ is the gyromagnetic ratio and $\lambda$ is the Gilbert damping factor. $\boldsymbol{M}$ is the surface magnetic moment with a magnitude of about a Bohr magnetic moment $|\boldsymbol{M}| = \mu_B$. The spin-exchange and spin-torque interaction around the moment introduce an effective magnetic field $|\boldsymbol{M}|\boldsymbol{H}_{\mathbf{eff}} = \alpha_M 2S_z/\hbar \boldsymbol{e}_{\boldsymbol{z}}$,[15] where $\alpha_M$ represents the interacting strength. The unit vector $\boldsymbol{e}_{\boldsymbol{z}}$ points to the left perpendicular to the substrate surface, and also defines the direction of the chiral axis. $S_z = \sum_{m=2L/3+1}^{2L} \frac{\hbar}{2}(n_m^{\uparrow} - n_m^{\downarrow})$ is the total spin accumulation of the sites close to $\boldsymbol{M}$ ($2L/3 < m \leq 2L$).

As shown in Figure 3, the spin accumulation $S_z$ ultimately decays to zero, and its time-averaged value during the precession remains finite. This finite average $S_z$ biases the final orientation of $\boldsymbol{M}$ toward the direction dictated by $\langle S_z \rangle$. When $S_z = 0$, the dynamic equilibrium $\frac{\mathrm{d}\boldsymbol{M}}{\mathrm{d}\tau} = 0$ is reached: precession and damping of $\boldsymbol{M}$ cease, and the out-of-plane component $M_z$ stabilizes at a non-zero value. In some real experimental setups,[4] the ferromagnetic substrate surface hosts a high density of molecules. Because the substrates are ferromagnetic, the initially disordered magnetic domains are aligned into a uniform ordered phase via CISM. This alignment is stabilized through magnetic interactions, enabling the magnetized state to persist for several days.[4]

## S4. Model parameters

We use typical geometric parameters of AHPAs, the same as our previous work:[6] radius $R = 0.25$ nm, twist angle $\Delta\varphi = \pm 5\pi/9$ (respectively for AHPA-L and AHPA-D), helical pitch $\Delta h = 0.15$ nm, and $l_c = 0.09$ nm. The nearest hopping in the AHPA is on the order of $t_1 \approx 150$ meV and is set to be the unit of energy. We use $\varepsilon_{sub} = 0$, $t_{sub} = 4t_1$, $t_c = 1.2t_1$, and $t_1^S = 0.05t_1$ to meet the experimental value of SOC,[16] which is about a few meV. The unit of the time in the whole paper is $\hbar/t_1 \approx 4$ fs. The strength of the electric field is set to be $V = -t_1$ without loss of generality. This electric field can result in an electrostatic force on an electron that is close to the typical magnitude of the dispersion force of about 10 pN.

The dephasing rate in AHPAs is set to be $\Gamma_{AHPA} = 0.01t_1$, to simulate a dephasing time scale $\hbar/\Gamma_{AHPA} \approx 400$ fs, which is close to the recent study.[15] The second kind of experiments can be performed on GaN with a long spin lifetime, so we set $\Gamma_{sub} = 0$. The third kind of experiments is usually performed on the surface of Au/ferromagnetism, so we set $\Gamma_{sub} = \Gamma_{AHPA}$ without the loss of generality, and discuss the case $\Gamma_{sub} \neq \Gamma_{AHPA}$ in Section S5.2.

For simplicity, $\gamma$ in Eq. (S5) is set to be the gyromagnetic ratio of electrons, and $\lambda = \gamma/|\boldsymbol{M}|$, same as a recent study.[15] We set a small interaction between the magnetic moment and spins $\alpha_M = 0.1t_1$ on the order of 10 meV[15] and a temperature close to the room temperature $\mathcal{T} = 0.2t_1/k_B \approx 300$ K in the main text (Figs. 1-3). This value of $\alpha_M$ is widely adopted and validated in analogous magnetization models,[15,17–19] and can be estimated in the following way. By comparing the Lifshitz-Landau-Gilbert equation used in our theoretical calculations with that employed in spin-transfer torque experiments, the effective magnetic field generated by the spin current can be expressed as:[20,21] $|\boldsymbol{H}_{\mathbf{eff}}| = \hbar J_s/2eM_s d$, where $J_s$ is the spin current density, $M_s$ the saturation magnetization, $d \approx 1$ nm represents the thickness of the magnetic layer interacting with spin-polarized electrons. Following the experimental definition of spin current,[20,21] the magnitude of the spin current in our theoretical framework can be estimated by $J_s = 2eS_z/\hbar\tau_0 S$, where $\tau_0 \approx 50$ fs represents the electron transit time across the interface, and $S \approx 1$ Å$^2$ denotes the effective area of the atom at the molecule end-substrate interface. From these expressions, we can get $\alpha_M = \hbar\mu_B/2M_s\tau_0 S d$. Taking $\mu_0 M_s = 0.5$ T ($\mu_0$ is the vacuum permeability), which is close to the real value of Ni,[22] the parameter $\alpha_M$ is estimated to be $\alpha_M \approx 15.4$ meV. The value $\alpha_M = 0.1t_1 = 15$ meV employed in our calculations aligns well with this estimated value, validating the

consistency of our theoretical approach. We further discuss the influence of $\alpha_M$ and $\mathcal{T}$ on the CISM in Figure S8, and CISM is considerable over a wide range of parameters.

# S5. Parameter dependence

In this section, we extend our analysis to the influence of various parameters on CISM under more realistic conditions.

## S5.1. Parameters of substrate

In the second experiment, we set $\varepsilon_{sub} = 0$ in $H_{sub}$, which is equal to the chemical potential of the system $\mu_{mol} = 0$. Under this condition, the electron number in the substrate described is $N_{sub} = \sum_i 1/\{\exp[(\varepsilon_i - \mu_{mol})/k_B\mathcal{T}] + 1\} = 30$. If $\varepsilon_{sub}$ changes, (for example, $\varepsilon_{sub} \neq \mu_{mol} = 0$), the electron number in the substrate will change, but this does not influence the electron transfer in the molecule and the spin accumulation process. Figure S9a shows the time evolution of total spin accumulation $S_z$ of AHPA-D in the near-2DEG region (analogous to Figure 2c in the main text) for varying $\varepsilon_{sub}$ ($\mu_{mol} = 0$ always). The $S_z$ curves corresponding to different substrate electron numbers closely overlap, demonstrating that the electron number in the substrate has no impact on our theoretical framework or computational results.

Moreover, in the second and third experiments, the coupling Hamiltonian $H_c$ of molecule and substrate may incorporate SOC: $H_c = t_c c_L^\dagger c_{L+1} + i s_c c_L^\dagger \sigma_x c_{L+1} + h.c.$ , where the term involving $\sigma_x$ is the SOC with the SOC strength $s_c$. This SOC term could allow for spin-flip processes as electrons transfer from the molecule to the substrate. However, as shown in Figure S9b (second experiment) and Figure S10a (third experiment), the $S_z$ curves for different $s_c$ values (up to $0.4t_1$, nearly ten times the SOC strength within the molecule $t_1^S$) closely overlap, indicating that the SOC in $H_c$ has negligible influence on the CISM results.

In the case of ferromagnetic substrates, it is more practical to account for spin-dependent band splitting in the substrate near the molecular chemical potential. When accounting for such splitting, the Hamiltonian of the substrate is:

$$H_{sub} = \sum_{m=L+1}^{2L} m_x c_m^\dagger \sigma_x c_m + \sum_{m=L+1}^{2L-1} \left( t_{sub} c_m^\dagger c_{m+1} + h.c. \right) \tag{S6}$$

where $m_x$ is the splitting energy. However, such splitting has almost no effect on the electron transfer in the molecule and the magnetization process, because the spin polarization from CISS is at the z direction, while the substrate is initially magnetized in-plane (at the x direction). Figure S10b shows the time evolution of the out-of-plane magnetic moment component $M_z$ (analogous to Figure 3e in the main text) under varying spin splitting $m_x$ in the third experiment. The $M_z$ curves corresponding to different splitting energies exhibit a similar increasing trend, demonstrating that spin splitting in the ferromagnetic substrate has negligible impact on our theoretical framework or computational results.

## S5.2. Temperature and dephasing

In realistic situations, dephasing is likely temperature-dependent: as the temperature $\mathcal{T}$ rises, dephasing effects induced by lattice vibrations (e.g., phonon scattering) may alter $\Gamma_{AHPA}$, thereby modifying the CISS efficiency. For the case of the third experiment, in Figure S10c, we calculate $M_z$ at time $\tau = 400$ (approximately saturated) under independently varying $\mathcal{T}$ and dephasing rates $\Gamma_{AHPA}$ ($\Gamma_{sub} = \Gamma_{AHPA}$). In realistic physical systems, dephasing is typically positively correlated with temperature. When considering their combined effects on $M_z$, multiple scenarios may emerge depending on parameter regimes. To illustrate this, we analyze two simplified scenarios:

1. If dephasing is predominantly thermally activated, $\Gamma_{AHPA}$ follows the trend indicated by the yellow arrow in Figure S10c. As the temperature $\mathcal{T}$ increases, $M_z$ increases with temperature. This trend aligns with recent experimental observations of CISS enhancement at elevated temperatures.[23]
2. If significant dephasing mechanisms persist even at zero temperature, $\Gamma_{AHPA}$ follows the trend indicated by the green arrow in Figure S10c. As the temperature $\mathcal{T}$ increases, $M_z$ decreases with temperature. Such suppression of CISS at higher $\mathcal{T}$ has also been experimentally reported.[24]

In reality, the dephasing mechanisms in the bulk substrate may differ significantly from those in the molecule, leading to $\Gamma_{sub} \neq \Gamma_{AHPA}$. Figure S10d shows the time-dependent out-of-plane magnetic moment $M_z$ under varying $\Gamma_{sub}$ strengths while fixing $\Gamma_{AHPA} = 0.01t_1$. The results demonstrate that the $M_z$ curves for different $\Gamma_{sub}$ share a similar rising trend. Notably, as $\Gamma_{sub}$ increases, the saturation value of $M_z$ progressively grows and ultimately stabilizes near $0.8|\boldsymbol{M}|$. This enhancement arises because a larger $\Gamma_{sub}$ broadens the energy levels in the substrate, facilitating electron injection into the substrate and thereby amplifying the saturation magnetization.

## S5.3. Geometric parameters

Finally, the structure parameters also influence CISM. For the case of the third experiment, in Figure S11, we calculate the time evolution of the out-of-plane magnetic moment component $M_z$ under varying molecule length $L$ (Figure S11a) and pitch $\Delta h$ (Figure S11b). Figure S11a shows that the magnetic component $M_z$ significantly enhances with increasing molecular length. This enhancement arises from the extended spin-selective transport distance within the molecule, which promotes the spin polarization efficiency.[11] Figure S11b shows that when the helical pitch is increased, the long-range electron hopping is suppressed, leading to a reduction in spin polarization efficiency and the magnetic component $M_z$.[6]

## S6. Additional figures

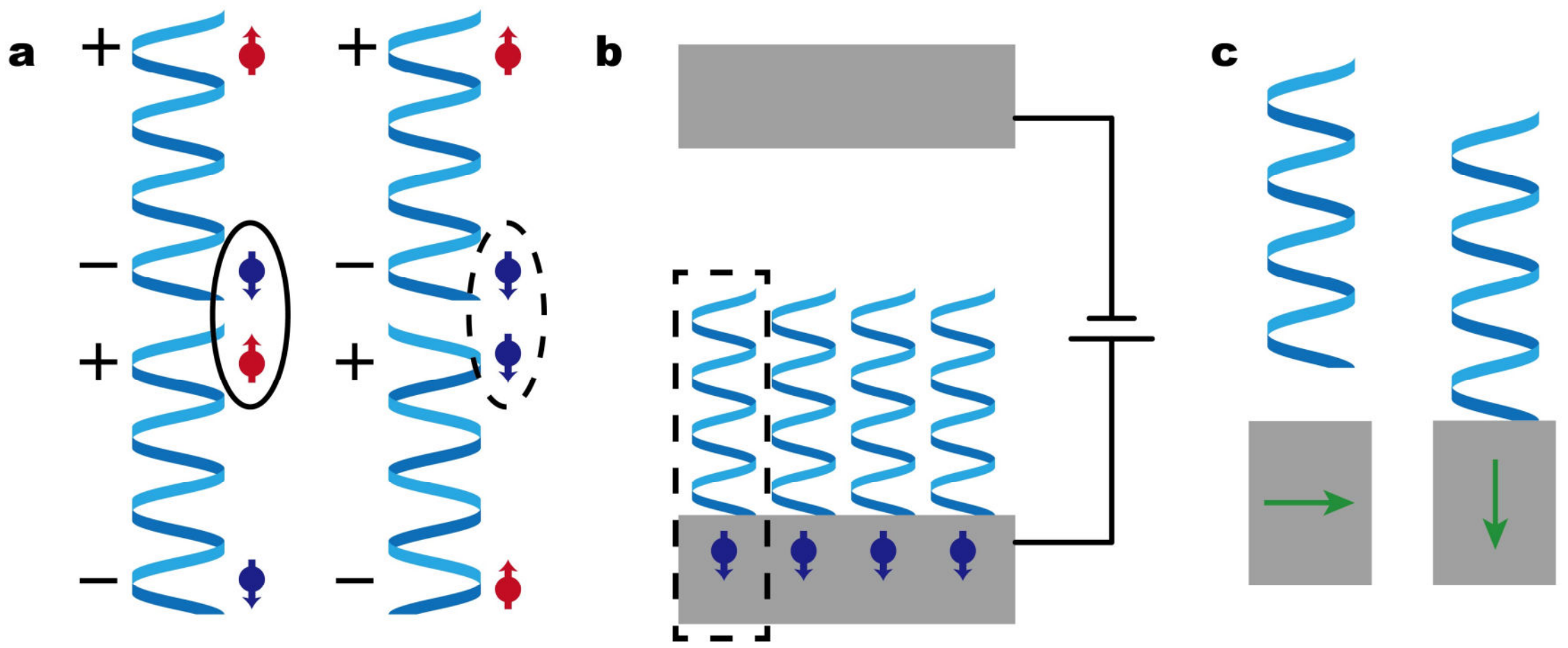

**Figure S1. The schematic of three types of chiral induced spin-polarization and magnetization.** (a) When two AHPAs interact with each other, the charge polarizes (represented by + and -), accompanied by spin-polarization (represented by red and blue arrows). Two AHPA-Ls (on the left) present a low-spin state with stronger coupling (shown as the solid box), while an AHPA-L and an AHPA-D (on the right) present a weak-coupling high-spin state (shown as the dashed box). (b) When AHPAs are driven by a gate voltage, AHE appears in the substrate as a signature of magnetization. The region enclosed by the dashed box corresponds to the upper panel of Figure 2a. (c) When the AHPA is self-assembled on a magnetic substrate with the magnetic moment initially in-plane magnetized, the direction of the magnetic moment will be twisted out of the plane, shown in green arrows.

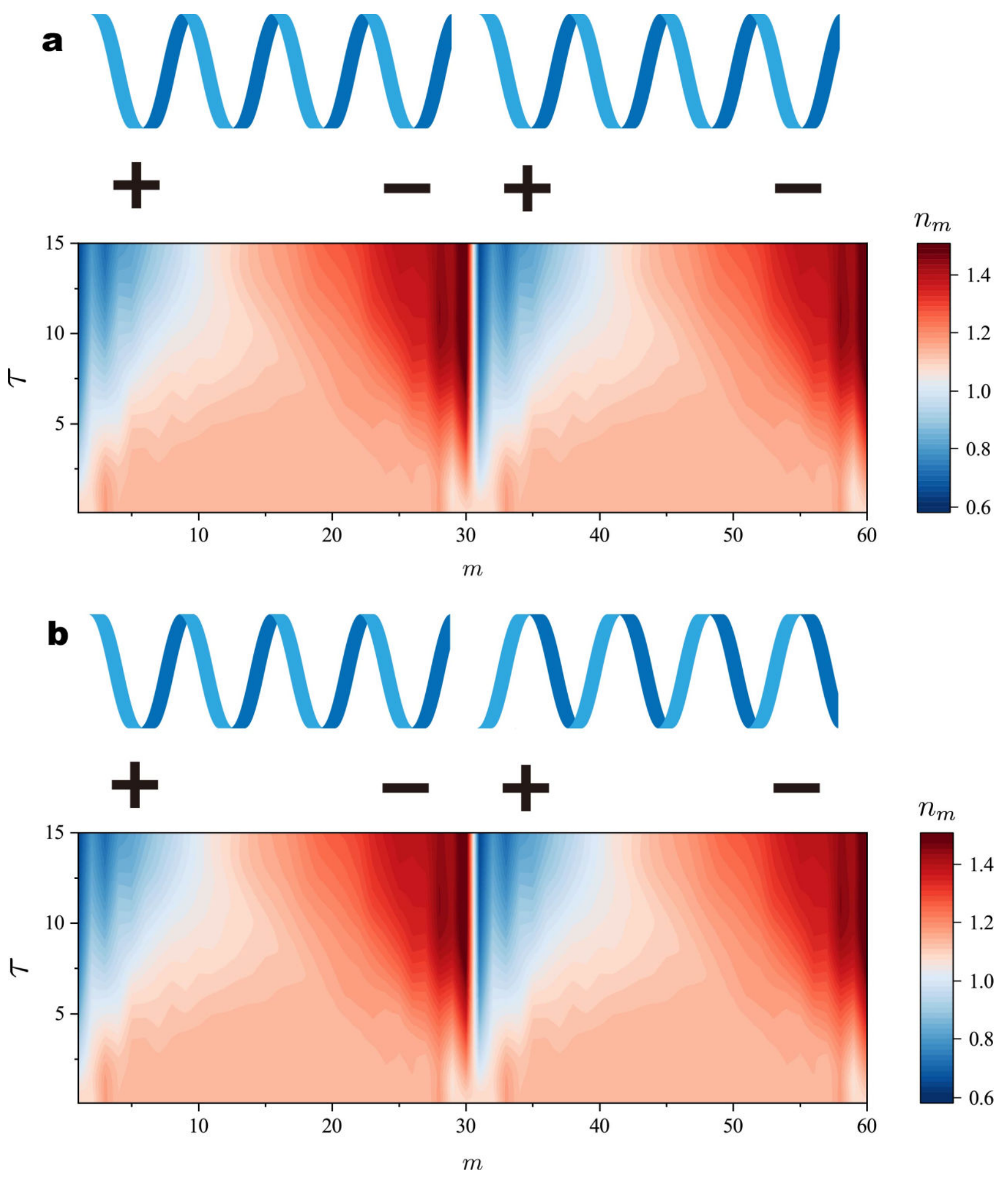

**Figure S2. The time evolution of the charge polarization of two AHPAs.** When two AHPAs interact with each other, the charge re-organizes, as a result of dispersion interaction. The schematics of the charge distribution of homochiral molecules (two AHPA-L) and heterochiral molecules (an AHPA-L and an AHPA-D) are shown in the upper panels of (a) and (b). Correspondingly, the time evolution of the electron density, represented by $n_m(\tau)$, is shown in the lower panels. The red region corresponds to a high electron density with charge "−", and the blue region corresponds to a low electron density with charge "+". The interface between the two AHPAs is located between $m = L = 30$ and $m = L + 1$. Both homochiral (**a**) and heterochiral (**b**) AHPAs exhibit almost identical charge polarization, with all AHPAs possessing electric dipole moments oriented in the same direction. The parameters used here are the same as those in Figure 1.

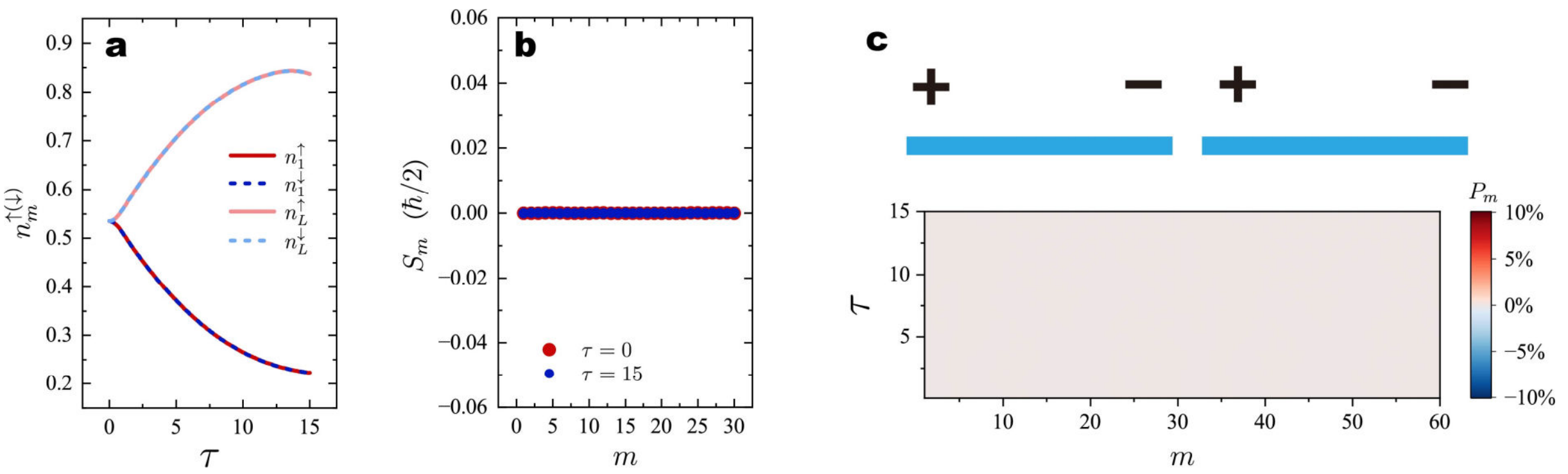

**Figure S3. The control group of two achiral molecules.** (a) The time evolution of the spin-resolved electron density at the first and the last site of an achiral molecule, represented by $n_1^{\uparrow(\downarrow)}$ and $n_L^{\uparrow(\downarrow)}$. The decrease in $n_1^{\uparrow(\downarrow)}$ and the increase in $n_L^{\uparrow(\downarrow)}$ indicate the presence of the charge-polarization. But $n_1^{\uparrow} = n_1^{\downarrow}$ and $n_L^{\uparrow} = n_L^{\downarrow}$ always hold true, showing that the electron density remains spin degenerate throughout the charge-polarization process. (b) The local spin accumulation $S_m$ of an achiral molecule at the time $\tau = 0$ and $\tau = 15$, which are always zeros due to the absence of the chirality. (c) Upper panel: The schematic diagram of two achiral molecules, with charge-polarization and without spin-polarization. Lower panel: The local spin polarization $P_m$ is zero for all sites and times, although the charge polarization is present in (a). The results in the lower panel correspond to the upper schematic diagram, where $m \leq L = 30$ ($L < m \leq 2L$) is the left (right) achiral molecule. The parameters used here are the same as those in Figure 1, except for the chirality ($\Delta\varphi = 0$ mentioned in the main text).

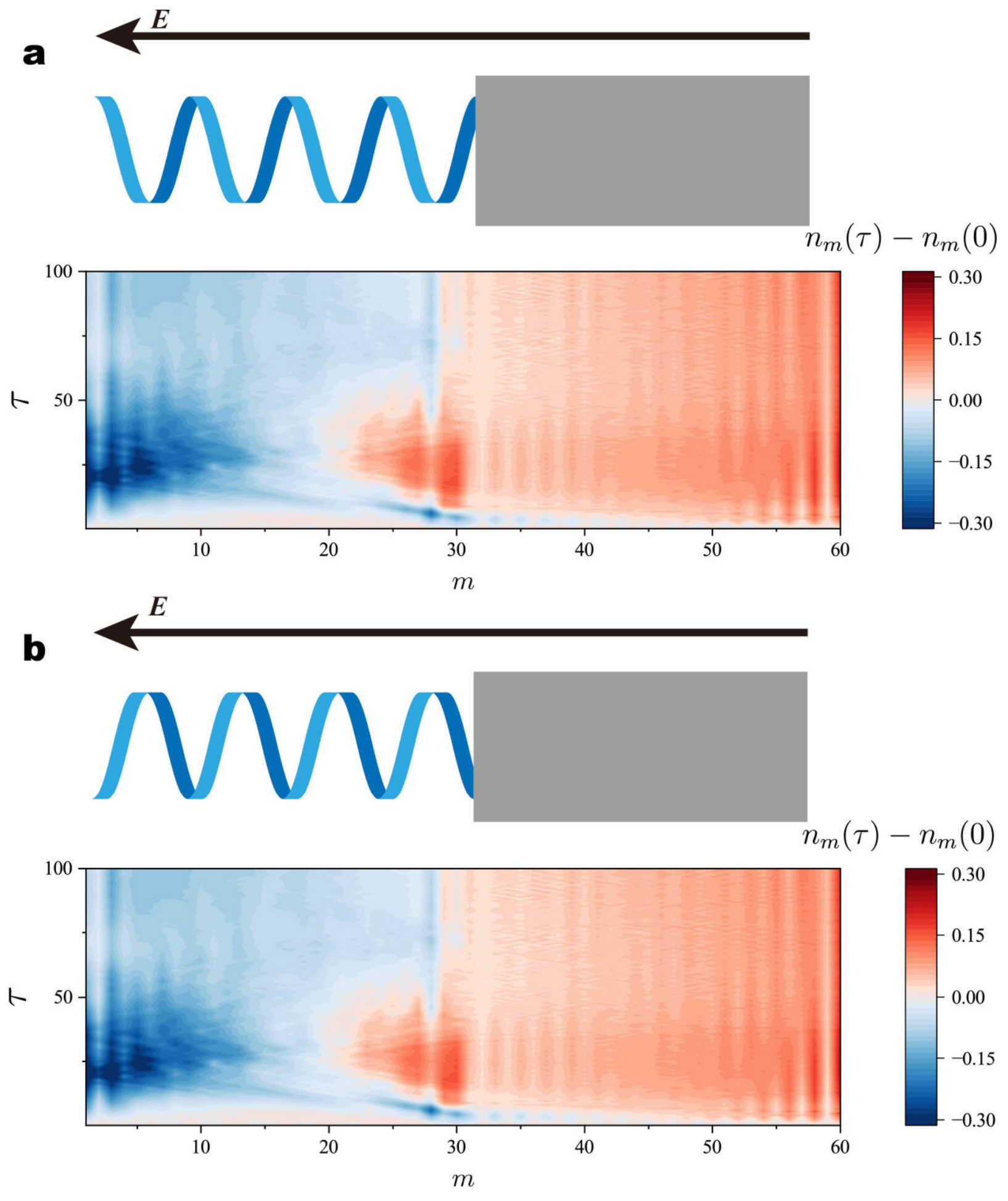

**Figure S4. The time evolution of the charge-polarization under an electric field of the AHPA/substrate system.** The upper panels in (a) and (b) are the schematics of an AHPA-L and an AHPA-D coupled to the substrate, respectively. The black arrows show the electric field introduced by the gate voltage. The lower panels in (a) and (b) show the time evolution of the charge-polarization for an AHPA-L/substrate system and an AHPA-D/substrate system, respectively, which is represented by the variation in electron density $n_m(\tau) - n_m(\tau = 0)$. The results in the lower panels correspond to the upper schematic diagrams, where $m \leq L = 30$ ($L < m \leq 2L$) is the AHPA (substrate). As the electric field is turned on, the substrate rapidly accumulates electrons (shown in red) and the molecules lose electrons (shown in blue), resulting in charge-polarization. The reflection of electrons at the system boundary results in some oscillations in the charge distribution with time. The distribution of charge is similar regardless of the chirality. The parameters used here are the same as those in Figure 2.

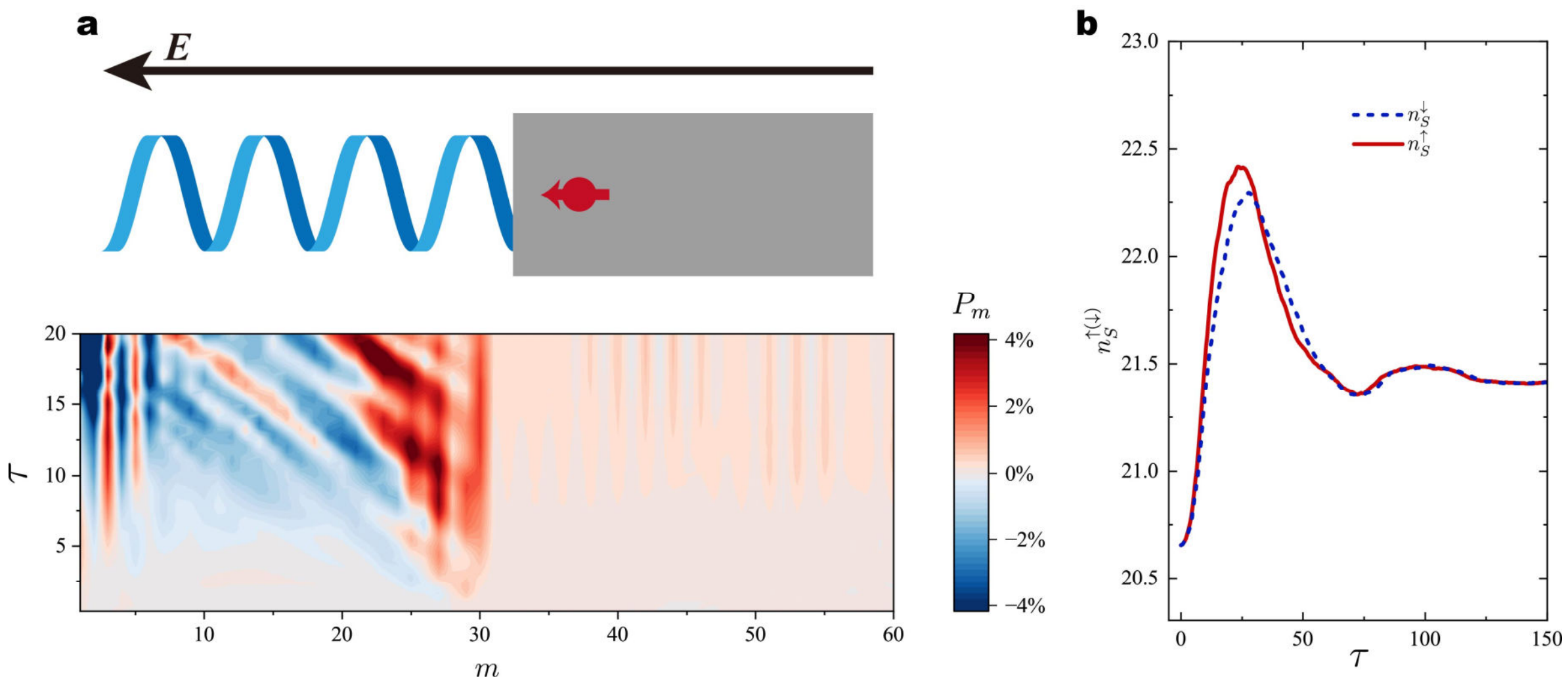

**Figure S5. The spin polarization under an electric field of the AHPA-D/substrate system. a** Upper panel: The schematic of an AHPA-D coupled to a substrate (the grey rectangle). The long black arrow shows the electric field and the short red arrow shows the polarized spin. Lower panel: The time evolution of the local spin polarization $P_m$ of the AHPA-D/substrate system under an electric field, which shows an opposite behavior to Figure 2a, as the chirality changed. (b) is the time evolution of $n_S^{\uparrow(\downarrow)}$ for the AHPA-D/substrate system, which reflects the spin-resolved number of electrons in the near-2DEG region with $2L/3 < m \leq 2L$. Similar to Figure 2b, the increase, oscillation, and stabilization at a higher than the initial value of the number of electrons $n_S^{\uparrow(\downarrow)}$ show the process of charge polarization. Due to the change of chirality, compared to Figure 2b, the relative magnitudes of $n_S^{\uparrow}$ and $n_S^{\downarrow}$ are reversed, where the originally larger component is now smaller, and vice versa. The parameters used here are the same as those in Figure 2, except for the chirality ($\Delta\varphi = -5\pi/9$ mentioned in the main text).

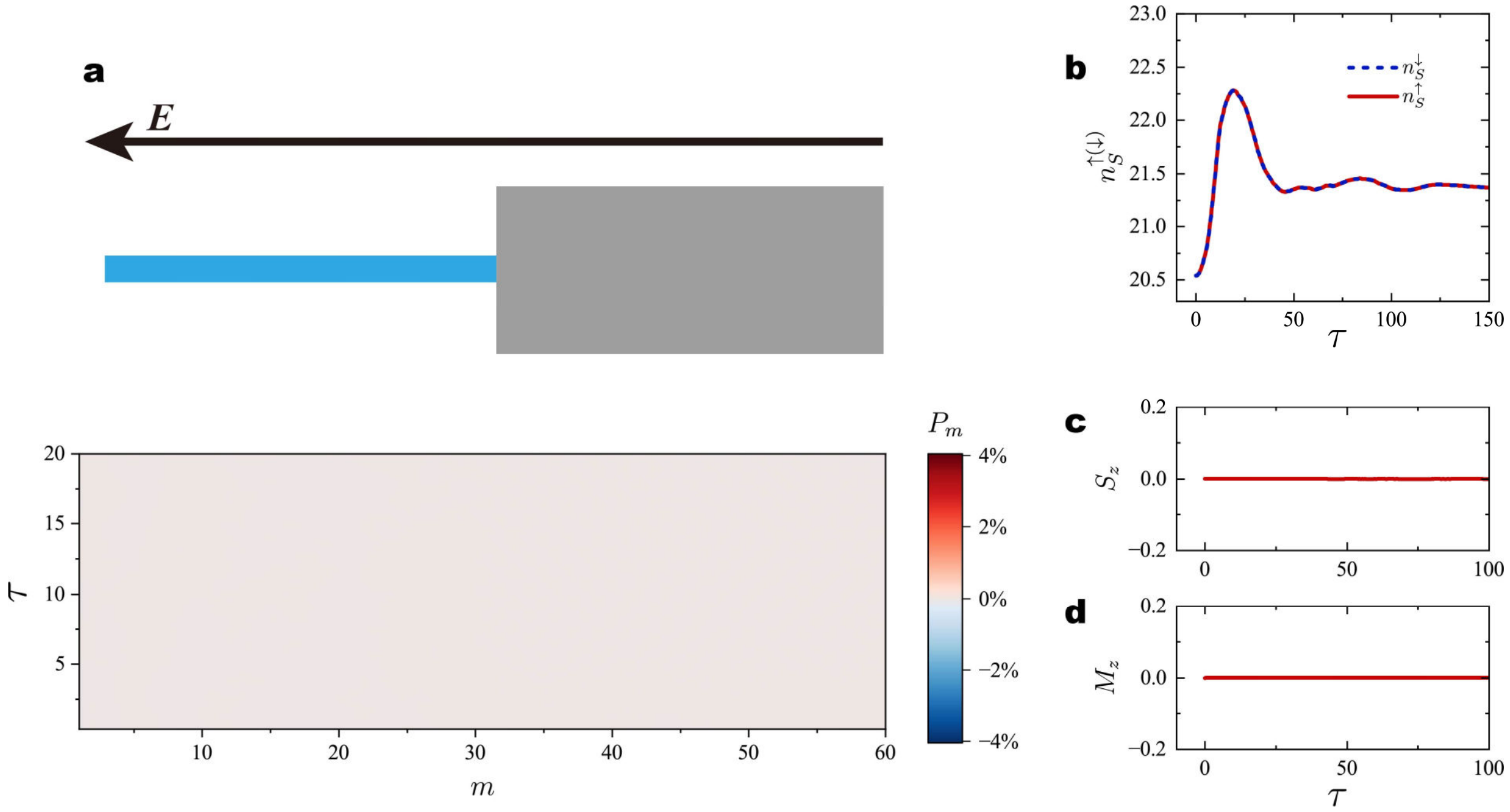

**Figure S6. The control group of the achiral molecule/substrate system under an electric field.** (a) Upper panel: The schematic diagram of an achiral molecule and its substrate under an electric field. Lower panel: The time evolution of the local spin-polarization $P_m$ of the achiral molecule/substrate system under an electric field. The results in the lower panel correspond to the upper schematic diagram, where $m \leq L = 30$ ($L < m \leq 2L$) is the achiral molecule (substrate). (b) The spin-resolved number of electrons in the region $2L/3 < m \leq 2L$ is represented by $n_S^{\uparrow(\downarrow)}$. (c) shows the total spin accumulation $S_z$ of electrons in the region $2L/3 < m \leq 2L$ for the achiral molecule. If there is a magnetic moment in the substrate coupling to the spin polarization $S_z$, the time evolution of the out-of-plane component of the magnetic moment $M_z$ is shown in (d). The results in Figure S6 show that no spin-polarization and no magnetization occur without the effect of chirality, although a charge-polarization exists, which highlights the essential role of chirality. The parameters used here are the same as those in Figure 2, except for the chirality ($\Delta\varphi = 0$ mentioned in the main text).

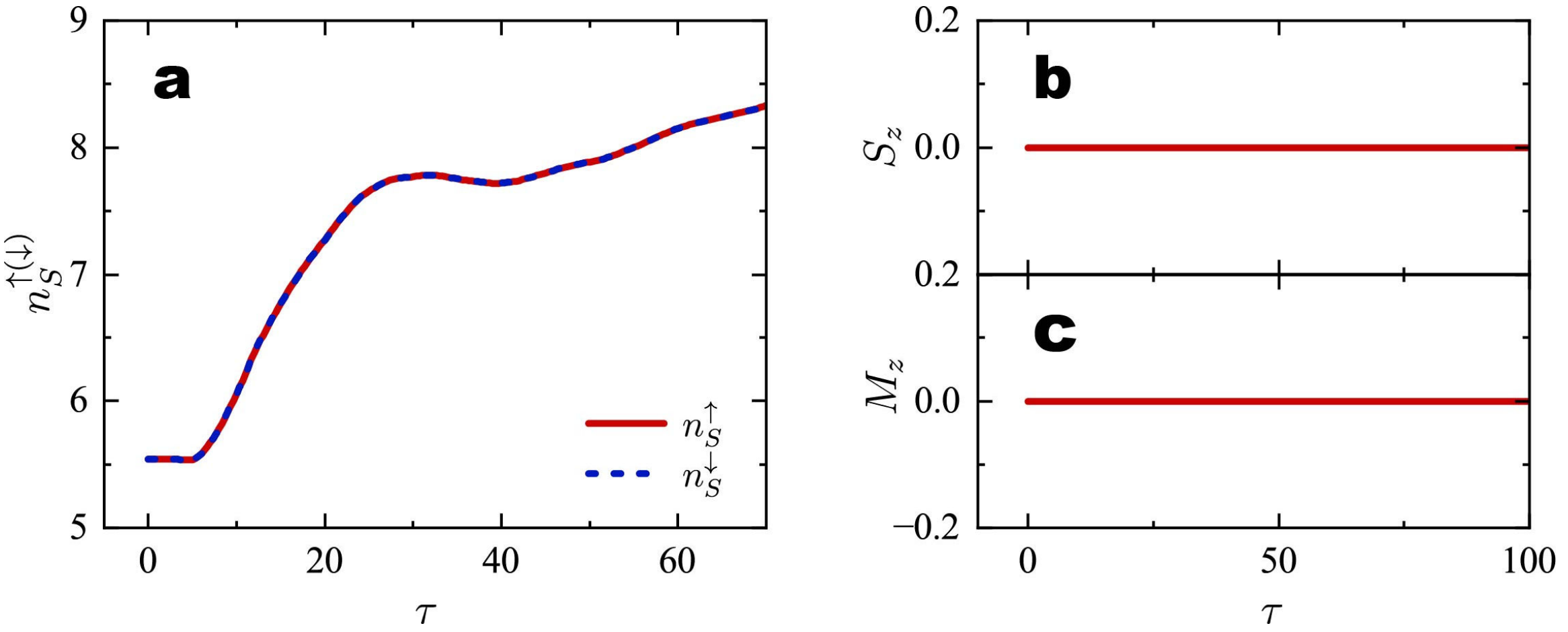

**Figure S7. The control group of the achiral molecule/ferromagnetic substrate.** (a) The time evolution of the spin-resolved number of electrons $n_S^{\uparrow(\downarrow)}$ in the substrate and near-substrate regions with $2L/3 < m \leq 2L$, after the molecule is adsorbed on the substrate. (b) shows the total spin accumulation of electrons $S_z$ in the region $2L/3 < m \leq 2L$ for the achiral molecule. (c) shows the time evolution of the out-of-plane component of the magnetic moment $M_z$. Compared Figure S8 to Figure 3, these results show that the chirality almost does not affect the charge transfer process. But without spin selection, the result becomes spin degenerate and without CISM, due to the achiral structure. The parameters used here are the same as those in Figure 3, except for the chirality ($\Delta\varphi = 0$ mentioned in the main text).

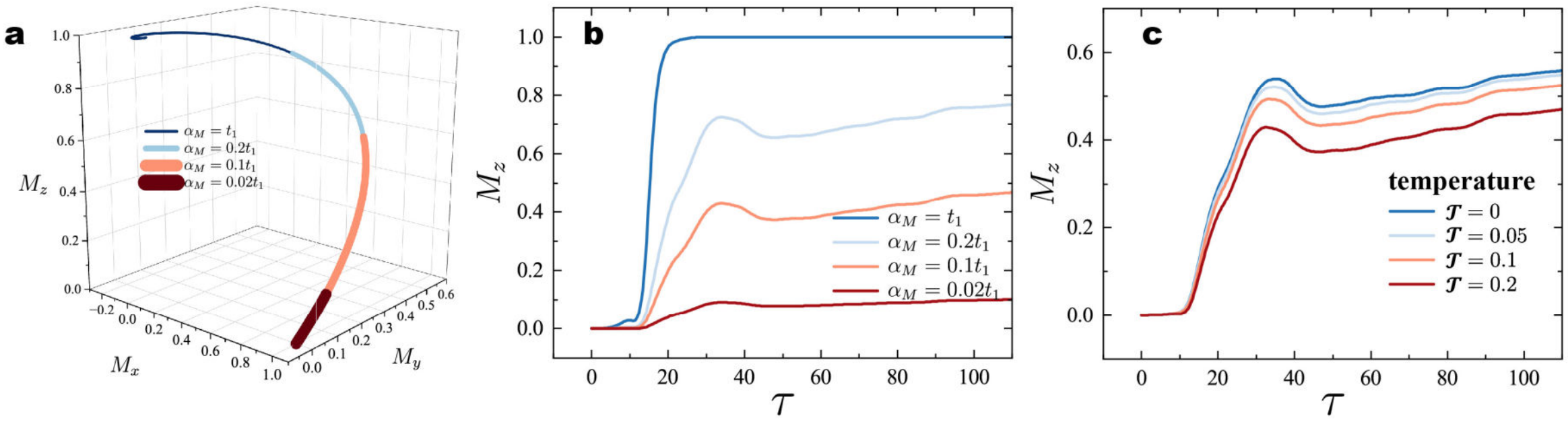

**Figure S8. The influence of parameters on CISM.** (a) Three-dimensional parametric plot of the trajectory of the magnetic moment $\boldsymbol{M}$ when the accumulated spin $S_z$ is coupled to $\boldsymbol{M}$ at different intensities $\alpha_M$ (mentioned in the main text). Without losing generality, the magnetic moment starts in the x direction. (b) The time evolution of $M_Z$ with different coupling strength $\alpha_M$. As the time $\tau$ increases, $\boldsymbol{M}$ experiences precession and damping, with $M_z$ increasing and saturating. A large $\alpha_M$ can facilitate the CISM, resulting in a large $M_z$. (c) The time evolution of $M_z$ with different temperature $\mathcal{T}$ (in the unit of $t_1/k_B$). In summary, CISM can be observed in a wide range of $\alpha_M$ (from $0.02t_1$ to $t_1$) and shows a robustness to $\mathcal{T}$ from zero to room temperature ($\mathcal{T} = 0.2$). We use AHPA-D here, and other parameters are the same as those in Figure 3.

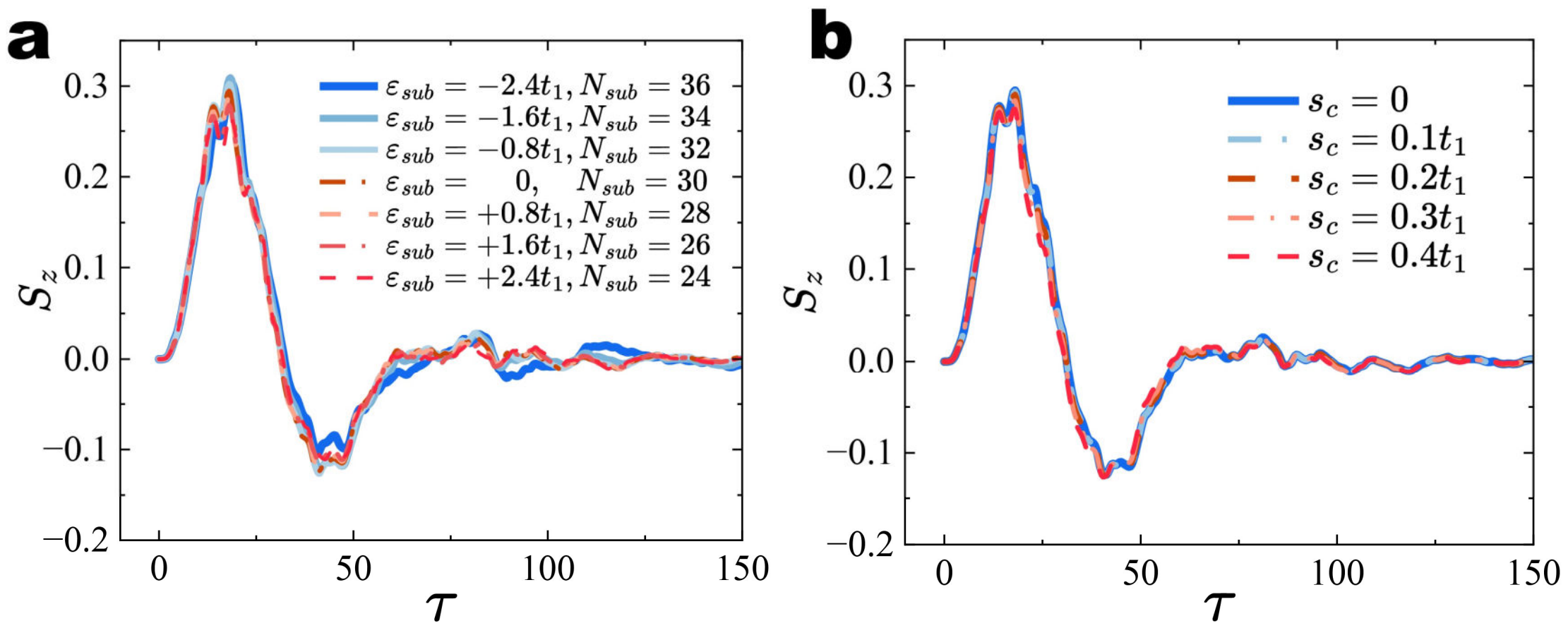

**Figure S9. The influence of the substrate on CISM in the second experiment.** The total spin accumulation $S_z$ in the near-2DEG region for varying the on-site energy of the substrate $\varepsilon_{sub}$(a) and SOC in the coupling Hamiltonian $s_c$ (b). The other parameters used here are the same as those in Figure 2.

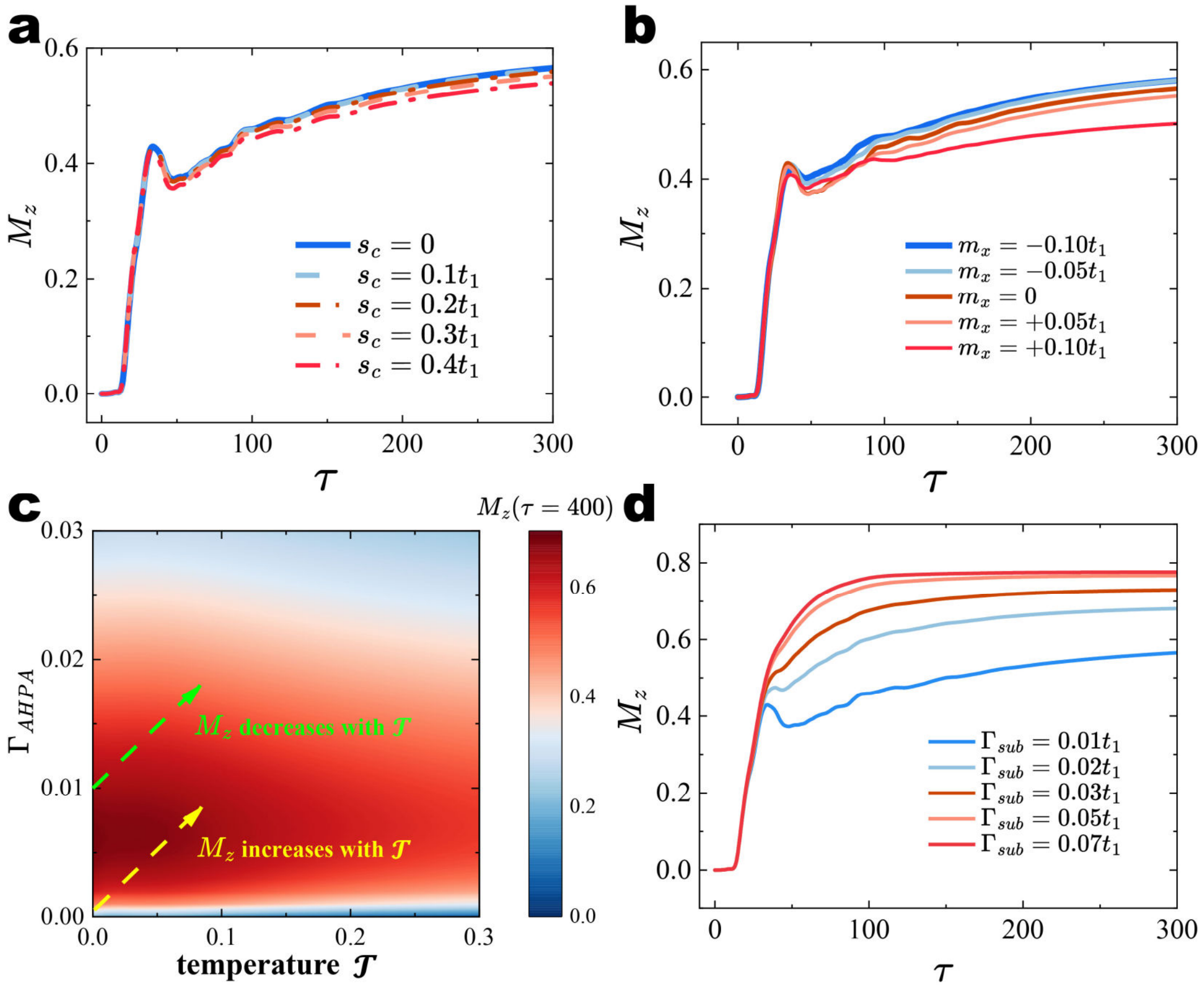

**Figure S10. The influence of the substrate, temperature, and dephasing on CISM in the third experiment.** The time evolution of the out-of-plane component of the magnetic moment $M_z$ for varying the SOC in the coupling Hamiltonian $s_c$ (a) and spin splitting energies $m_x$ (b). (c) $M_z$ at time $\tau = 400$ under varying temperature $\mathcal{T}$ and dephasing rates $\Gamma_{AHPA}$. The yellow and green arrows show two scenarios discussed in Section S5.2. (d) Time evolution of $M_z$ for varying $\Gamma_{sub}$ in the third experiment. The molecule here is AHPA-D, and other parameters used here are the same as those in Figure 3.

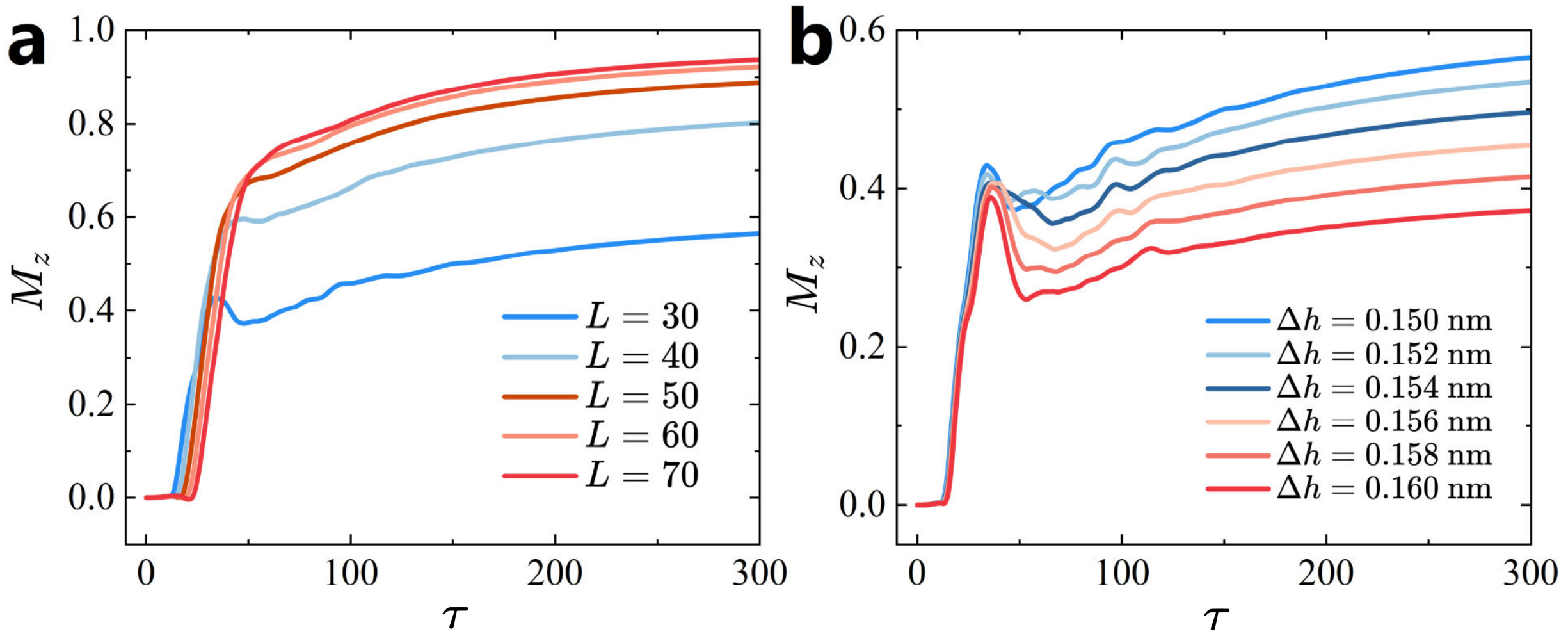

**Figure S11. The influence of geometric parameters on CISM in the third experiment.** In AHPA-D, the time evolution of the out-of-plane magnetic moment component $M_z$ is studied under varying molecule length $L$ (Figure S11a) and pitch $\Delta h$ (Figure S11b). Other parameters are the same as those in Figure 3.